\documentclass[12pt,english]{article}

\usepackage[letterpaper]{geometry}
\usepackage[font=small, width=0.9\textwidth]{caption}

\usepackage[T1]{fontenc}
\usepackage[latin9]{inputenc}

\usepackage{amsmath}
\usepackage{amsthm}
\usepackage{amssymb}
\usepackage{authblk}
\usepackage{babel}
\usepackage{graphicx}
\usepackage{hyperref}
\usepackage{lmodern}
\usepackage{mathrsfs}
\usepackage{sistyle}
\usepackage{subfigure}  
\usepackage{verbatim}   
\usepackage{upgreek}
\makeatletter

\usepackage[numbers,sort&compress,comma]{natbib}

\usepackage{hyperref}
\hypersetup{
	colorlinks=true,
	linkcolor=blue,
	urlcolor=red,
	citecolor=blue
}

\usepackage{fancyhdr}
\usepackage{enumitem}
\setenumerate{itemsep=5pt,topsep=5pt}

\renewcommand\section{%
	\@startsection {section}{1}%
	{\z@}{-16pt \@plus 0pt \@minus 0pt}%
	{8pt \@plus.2ex}%
	{\normalfont\large\bfseries\scshape}}

\renewcommand\subsection{%
	\@startsection {subsection}{1}%
	{\z@}{-8pt \@plus 0pt \@minus 0pt}%
	{4pt \@plus.2ex}%
	{\normalfont\normalsize\bfseries}}

\renewcommand\subsubsection{%
	\@startsection {subsubsection}{1}%
	{\z@}{-8pt}%
	{0.001pt \@plus 0pt}%
	{\normalfont\normalsize\bfseries}}

\renewcommand{\paragraph}{%
	\@startsection{paragraph}{4}%
	{\z@}{-8pt}{-0.5em}%
	{\normalfont\normalsize\bfseries}%
}

\makeatother

\newcommand{\um}{\upmu\text{m}}

\newcommand{\degC}{$^\circ$C}
\newcommand{\SiN}{Si$_3$N$_4$}

\begin{document}
	
\title{Flexure-Tuned Membrane-at-the-Edge Optomechanical System}

\author{Vincent Dumont\footnote{vincent.dumont@mail.mcgill.ca}, Simon Bernard, Christoph Reinhardt, Alex Kato,  \\
	Maximilian Ruf, and Jack C. Sankey\footnote{jack.sankey@mcgill.ca}}
\affil{McGill University Department of Physics\\
	 3600 rue University, Montr\'{e}al QC, H3A 2T8, Canada}

\maketitle



\begin{abstract}
\noindent
We introduce a passively-aligned, flexure-tuned cavity optomechanical system in which a membrane is positioned microns from one end mirror of a Fabry-Perot optical cavity. By displacing the membrane through gentle flexure of its silicon supporting frame (i.e., to $\sim$80~m radius of curvature (ROC)), we gain access to the full range of available optomechanical couplings, finding also that the optical spectrum exhibits none of the abrupt discontinuities normally found in ``membrane-in-the-middle'' (MIM) systems. More aggressive flexure (3 m ROC) enables $>$15 $\um$ membrane travel, milliradian tilt tuning, and a wavelength-scale ($1.64 \pm 0.78~\um$) membrane-mirror separation. We also provide a complete set of analytical expressions for this system's leading-order dispersive and dissipative optomechanical couplings. Notably, this system can potentially generate orders of magnitude larger linear dissipative or quadratic dispersive strong coupling parameters than is possible with a MIM system. Additionally, it can generate the same \emph{purely} quadratic dispersive coupling as a MIM system, but with significantly suppressed linear dissipative back-action (and force noise). 
\end{abstract}

\pagebreak
\tableofcontents

\pagebreak

\section{Introduction}

In the field of optomechanics, radiation forces have provided previously inaccessible control over mechanical objects of all sizes, with systems increasingly often exhibiting quantum properties of motion and light \cite{Aspelmeyer2014Cavity}. Within this context, the so-called ``membrane-in-the-middle'' (MIM) geometries \cite{Thompson2008Strong, Bhattacharya2008Optomechanical} have enjoyed great success, owing in large part to (i) separating the tasks of fabricating high quality mirrors and low-noise mechanical elements, and (ii) the exquisite properties of \SiN, which is capable of achieving extremely low force noise \cite{Ghadimi2018Elastic, Reinhardt2016Ultralow, Norte2016Mechanical}, high $Q\times f$ products \cite{Tsaturyan2017Ultracoherent, Ghadimi2018Elastic}, and optical losses compatible with ultrahigh cavity finesse $\sim 10^6$ \cite{Wilson2009Cavity, Sankey2010Strong, Reinhardt2016Ultralow}. To date, MIM systems exhibiting linear dispersive coupling (wherein the membrane's displacement modulates the cavity's resonant frequency \emph{linearly}) have been used to laser-cool millimeter-scale membranes to the quantum realm \cite{Purdy2015Optomechanical, Underwood2015Measurement} and the back-action limit \cite{Peterson2016Laser}, observe quantum radiation pressure noise \cite{Purdy2013Observation}, squeeze light \cite{Nielsen2017Multimode, Purdy2013Strong}, approach the standard quantum limit \cite{Rossi2018Measurement}, and transfer information between microwave and optical carriers \cite{Andrews2014Bidirectional, Higginbotham2018Harnessing}.

Since the membrane inherently interacts with \emph{two} cavity modes, it is also possible to generate purely \emph{quadratic} dispersive coupling \cite{Thompson2008Strong, Sankey2010Strong}, providing access to stable optical springs \cite{Lee2014Multimode} applicable to enhancing mechanical $Q$-factors \cite{Ni2012Enhancement, Chang2012Ultrahigh, Muller2015Enhanced} or tuning the shape of a mechanical mode \cite{Barasheed2016Optically}, and unique squeezing opportunities \cite{Nunnenkamp2010Cooling}. Additionally, the existence of purely quadratic dispersive coupling suggests the possibility of quantum nondemolition (QND) readout of the membrane's phonon number states \cite{Thompson2008Strong, Jayich2008Dispersive}. However, even with a lossless single-port cavity, the linear \emph{dissipative} coupling (wherein the membrane displacement modulates the cavity's decay rate linearly) imposes the yet-illusive requirement of single-photon strong coupling to resolve an energy eigenstate before it is demolished \cite{Miao2009Standard,Yanay2016Quantum}.  On the other hand, such dissipative coupling could enable ground-state cooling in the ``fast-cavity'' limit \cite{Elste2009Quantum}, provide more squeezing opportunities \cite{Kilda2016Squeezed}, and generate stable optical springs (studied thus far in the Michelson-Sagnac geometry \cite{Tarabrin2013Anomalous, Sawadsky2015Observation}). 

Finally, fiber-coupled micro-mirrors have enabled MIM systems to achieve orders of magnitude stronger optomechanical coupling by reducing the cavity length from centimeters to $\sim 80~\um$ \cite{Flowers2012Fiber}; one outstanding goal is to create a stable, wavelength-scale MIM cavity, thereby achieving a per-photon force comparable to that normally associated with nanoscale photonic/phononic crystals -- which themselves have achieved impressive results (e.g., \cite{Aspelmeyer2014Cavity,  Arrangoiz2019ResolvingARXIV, Meenehan2015Pulsed, Marinkovic2018Optomechanical, Riedinger2018Remote, Hong2017Hanbury}) -- though deep within the fast-cavity limit. Wavelength-scale \emph{membrane-membrane} separations (currently limited to $\sim 8$~$\um$ \cite{Nair2017Optomechanical}) are also desirable for nested cavity systems, wherein the optomechanical coupling can be enhanced  \cite{Piergentili2018Two}, potentially even to the single-photon strong coupling limit \cite{Li2016Cavity}.

Here, we present a simple geometry in which a membrane is positioned very close to one end mirror of a Fabry-Perot cavity. In Sec.~\ref{sec:optomechanics}, we derive analytical expressions for the leading-order optomechanical couplings, notably finding that this ``membrane-at-the-edge'' (MATE) geometry could be used to alleviate the requirement of single-photon strong coupling for QND readout, or to generate larger optomechanical couplings. We then demonstrate and characterize a simple, passively aligned, flexure-tuned MATE system. As discussed in Sec.~\ref{sec:resonances}, gently flexing the membrane's silicon supporting frame (to 80 m radius of curvature (ROC)) allows us to smoothly tune the form of the optomechanical coupling through its full range. Advantageously, we observe a smooth variation in the cavity response over wavelength-scale membrane displacements, with none of the discontinuities normally found in MIM systems (consistent with our model). In Sec.~\ref{sec:wavelength-scale}, we show how aggressive flexure (3 m ROC) enables large ($>15~\um$) membrane travel and a mirror-membrane separation comparable to the laser wavelength. By applying pressure at a single point, we also tune the membrane's tilt (relative to the mirror surface) by 0.7 mrad over the full travel range while protecting the membrane from collision with the mirror. Importantly, this monolithic geometry poses significantly fewer technical challenges than alignment with multi-axis stages, and, for small ROC, reduces susceptibility to dust at small separations, as in convex lens induced confinement (CLIC) systems \cite{Leslie2010Convex}. In addition to the aforementioned advantages, this work paves the way toward flexure-tuned fiber-coupled Fabry-Perot MIM systems and membrane-membrane cavities at the wavelength scale.

\begin{figure}[!ht]
	\centering
	\includegraphics[width=10cm]{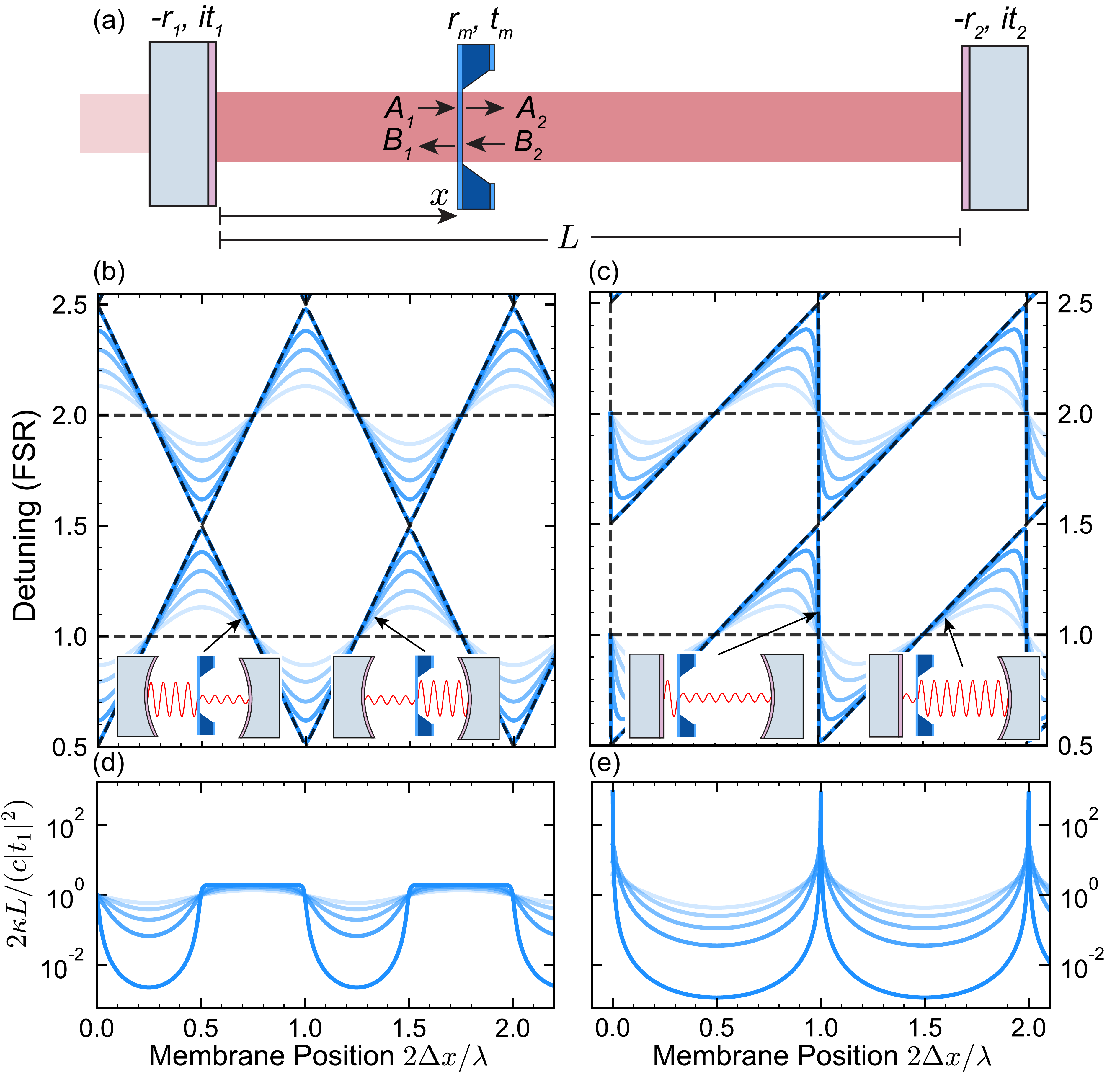}
	\caption{Optical resonances for a membrane in a cavity. (a) 1D model, comprising two fixed end mirrors (gray), with reflection (transmission) coefficients $-r_j$ ($it_j$), and a thin, flexible dielectric membrane (blue) having reflection (transmission) coefficients $r_\text{m}$ ($t_\text{m}$). $A_j$ ($B_j$) indicate the right-moving (left-moving) field amplitudes just outside the membrane. (b)-(c) Dependence of the cavity resonance's detuning (normalized by the free spectral range (FSR)) on membrane displacement $\Delta x$ from (b) the cavity center ($x$=$L/2$) and (c) the first mirror ($x$=0), with $r_\text{m}$ values (from light to dark) -0.4, -0.6, -0.8, -0.931, and -0.9977 ($r_\text{m}$'s phase chosen to highlight avoided crossings as in \cite{Thompson2008Strong}). Horizontal dashed lines represent empty cavity resonant frequencies ($r_\text{m}$=0), while other dashed lines represent the left (negatively sloped) and right (positively sloped) sub-cavity resonances when $t_\text{m}$=0. Insets qualitatively show the field distribution of these modes. (d)-(e) Dependence of cavity's energy decay rate $\kappa$ on membrane displacement $\Delta x$ from (d) the cavity center and (e) the first mirror, normalized by the empty cavity value for a single-port cavity ($t_2$=0, $r_2$=-1).}
	\label{fig1}
\end{figure}

\section{Optomechanics with a membrane at the edge}\label{sec:optomechanics}

A MATE system exhibits both quantitative and qualitative differences from a MIM system, the most essential of which are captured by a 1D scattering model drawn in Fig. \ref{fig1}(a). Extending the formalism of \cite{Jayich2008Dispersive} (Appendix~\ref{app:scattering}), we consider a cavity comprising two fixed end mirrors (gray) separated by a length $L$, with reflection (transmission) coefficients $-r_j$ ($it_j$). A membrane (blue) having reflection (transmission) coefficients $r_\text{m}$ ($t_\text{m}$) is positioned in between. For high-finesse cavities ($r_j\rightarrow 1$), the right-moving ($A_j$) and left-moving ($B_j$) fields of each sub-cavity differ only by a phase, and, for long cavities $L\gg \lambda$ (where $\lambda=2\pi/k$ is the light's wavelength and $k$ is its wavenumber), the cavity's resonant frequencies are
\begin{equation}\label{eq:intro-omega-mim}
\omega_\text{MIM} \approx ck_\text{MIM} =  N\omega_\text{FSR}+\frac{\omega_\text{FSR}}{\pi}\left(\arccos\left[(-1)^{N+1}|r_\text{m}|\cos(2 k_N\Delta x)\right]- \phi_\text{r}\right)
\end{equation}
for a MIM geometry (small displacements $\Delta x$ from the cavity center), and
\begin{align}\label{eq:intro-omega-mate}
\omega_\text{MATE} \approx c k_\text{MATE} = N\omega_\text{FSR} + \frac{\omega_\text{FSR}}{\pi}\arctan\left[\frac{\cos(\phi_\text{r})+|r_\text{m}|\cos(2k_N \Delta x)}{\sin(\phi_\text{r})-|r_\text{m}|\sin(2k_N \Delta x)}\right],
\end{align}
for a MATE geometry (small displacements $\Delta x$ from either end mirror), where $k_\text{MIM}$ and $k_\text{MATE}$ are the resonant wavenumbers, $\omega_\text{FSR}=\pi c/L$ is the empty cavity's free spectral range, $\phi_\mathrm{r}$ is the phase change for light reflected from the membrane, and $k_N=\pi N/L$ is the $N^\text{th}$ empty-cavity resonance, with $N\gg 1$. These frequencies are plotted in Fig.~\ref{fig1}(b) for the MIM system and Fig.~\ref{fig1}(c) for the MATE system. In both cases, the membrane divides the cavity into left and right sub-cavity modes (each having resonant frequencies indicated by black dashed lines) and hybridizes them via its transmission. This produces the shown networks of avoided crossings that define the dispersive optomechanical coupling. 

At the same time, the $x$-dependence of the power landing on each end mirror (see inset sketches in Fig. \ref{fig1}~(b)-(c)) leads to an $x$-dependent cavity decay rate
\begin{align}\label{eq:intro-kappa}
\kappa = \frac{(1-|r_\text{m}|^{2})c|t_1|^2+(1+2|r_\text{m}|\cos(2kx+\phi_\text{r})+|r_\text{m}|^{2})c|t_2|^2} {2x(1-|r_\text{m}|^{2})+2(L-x)(1+2|r_\text{m}|\cos(2kx+\phi_\text{r})+|r_\text{m}|^{2})}.
\end{align}
for any location $x$ in the cavity, where $k$ (here) is the resonant wavenumber. This expression is plotted for the ``single-port'' ($|r_1|<1$, $r_2\rightarrow-1$) case in Fig.~\ref{fig1}(d) with $k=k_\text{MIM}$, and in Fig.~\ref{fig1}(e) with $k=k_\text{MATE}$ and the membrane positioned a distance $\Delta x \ll L$ from the first ``input'' mirror. This highlights the aforementioned linear dissipative coupling (non-zero $\partial_x\kappa$) occurring at positions where the dispersive coupling is purely quadratic. As discussed in Appendix~\ref{app:scattering}, however, $\partial_x\kappa$ at these locations can be suppressed by a factor $2\Delta x / L\ll 1$ simply by positioning the membrane a distance $\Delta x$ from the second ``backstop'' mirror. This similarly suppresses the added radiation force noise, thereby relaxing the requirement of single-photon strong coupling for QND phonon number readout (the subject of a forthcoming quantum treatment); this suppression can also benefit the pursuit of low-noise optically trapped (levitated) systems exploiting quadratic dispersive coupling. On the other hand, the linear dissipative ``strong coupling'' parameter $(\partial_x\kappa)x_\text{zpf}/\kappa$ (a unitless figure of merit that, when greater than one, indicates the strong coupling regime, where $x_\text{zpf}$ is the membrane's zero-point motion) is twice that of a MIM system when the membrane is instead positioned near the first ``input'' mirror.

The large asymmetry of the sub-cavities enables two additional enhancements. First, while the value $\partial_x^2\omega$ at the extrema of $\omega_\text{MATE}$ is identical to that of the MIM system, moving the membrane away from the extrema (toward the steep side of the sawtooth in Fig.~\ref{fig1}(c)) provides an increase in $\partial_x^2\omega$ by a factor as large as $\sim 1/|t_m|^3$ in the low-$|t_m|$ limit. This comes at the expense of increased $\kappa$, but nonetheless corresponds to an increase in the corresponding strong coupling parameter $(\partial_x^2 \omega) x_\text{zpf}^2 / \kappa$ by $\sim 1/|t_m|$. Second, again by moving the membrane away from the extrema, the \emph{dissipative} strong coupling parameter $(\partial_x \kappa)x_\text{zpf}/\kappa$ can be similarly increased by a factor $\sim 1/|t_m|$. Note these enhancements are most relevant for cavities incorporating a high-reflectivity mechanical element, such as a structured membrane \cite{Stambaugh2015From,Chen2017High} (unpatterned SiN structures can only produce few percent enhancement, while higher-index materials (e.g., Si) can achieve $\sim$50\%), but the same formalism is relevant for coupled transverse modes \cite{Sankey2010Strong} having very different linear couplings and nearfield systems on chip \cite{Paraiso2015Position}.

Finally, as a practical concern, we emphasize that the $x$-dependence of $\omega_\text{MATE}$ is approximately identical for adjacent optical modes, meaning the \emph{full} spectrum of cavity modes should exhibit no abrupt avoided crossings with higher-order transverse modes. This will yield consistent cavity performance over a large range of $\Delta x$, easing the tasks of translating the membrane while locked (e.g., to tune the optomechanical coupling) and / or tracking large-amplitude vibrations.

\section{Optical resonances of flexural MATE system}\label{sec:resonances}

\begin{figure}[!ht]
	\centering
	\includegraphics[width=10cm]{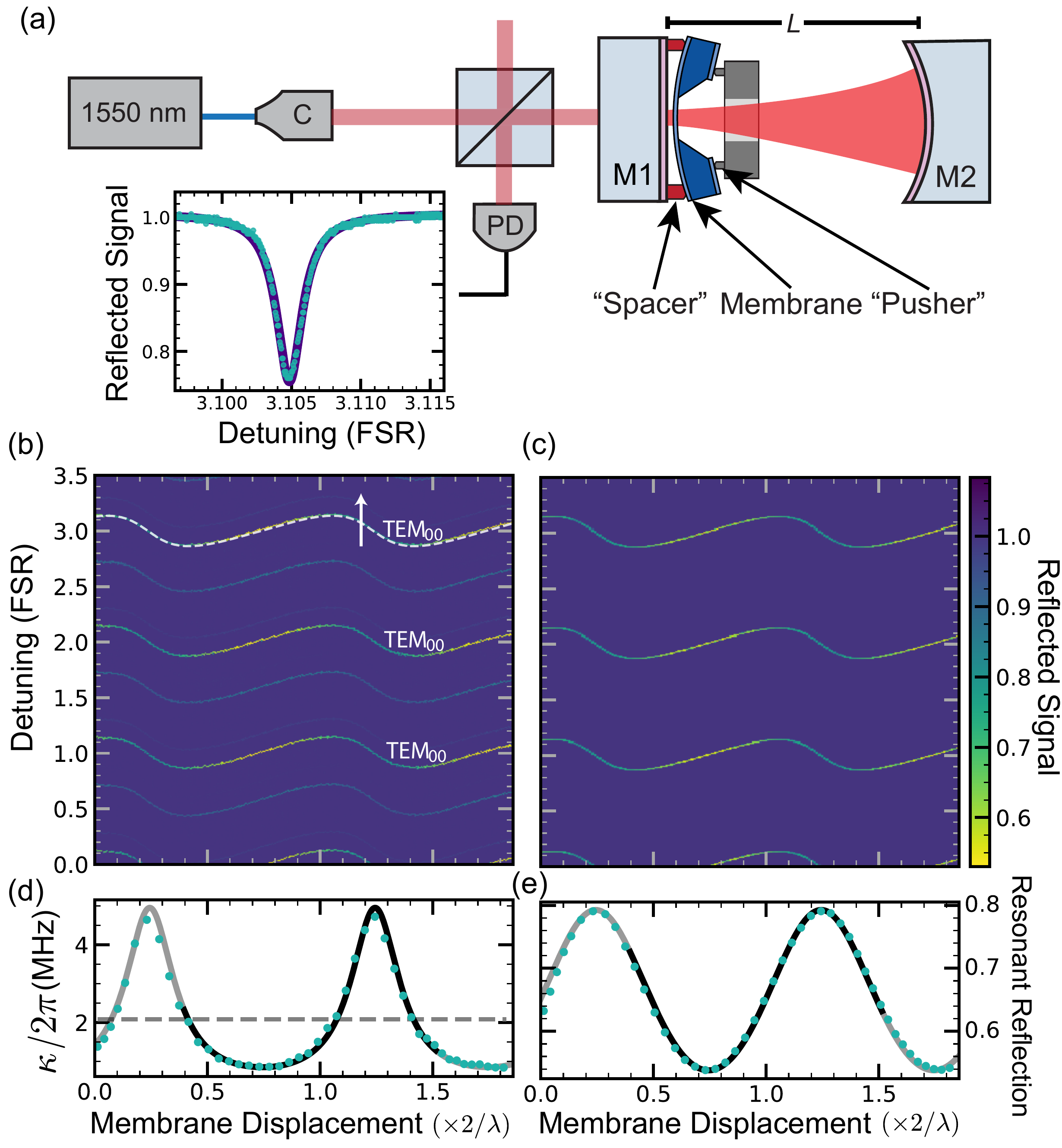}
	\caption{Optical response of flexural MATE system. (a) Diagram of measurement system and MATE cavity (not to scale), comprising an input mirror M1, $21~\um$-thick spacers, a membrane, piezo-actuated pushers, and a ``backstop'' mirror M2. 10 mW of laser light ($\lambda=1550$~nm) from a fiber collimator C passes through a 50:50 splitter, is reflected from M1, and collected with a photodiode PD. The length $L = 10$ cm between M1 and M2 is tuned with piezo actuators on M2; the inset shows the fractional reflected power (normalized by the off-resonance value 2.5 mW) as a function of the laser's detuning from an $L$-swept cavity resonance. The violet curve is a ``typical'' Lorentzian fit used to extract the cavity's energy decay rate $\kappa$. (b) Same measurement for a range of $L$ at varied membrane displacements $\Delta x$. The white dashed line shows part of a simultaneous fit to three modes' resonant frequencies (Eq.~\ref{eq:intro-omega-mate}) used to eliminate piezo nonlinearities, and the arrow shows the location and direction of the sweep in (a). Higher-order transverse modes appear as faint resonances. (c) Spectrum predicted by a 1D transfer matrix model, with parameters estimated as described in the main text. (d) Dependence of cavity decay rate $\kappa$ on membrane position for the central TEM$_{00}$ (brightest) resonance in (b). The solid line is $\kappa(x)$ obtained by numerically solving the 1D model, and the black part highlights the region used for our main fits (including the piezo corrections). The dashed line is the fit empty cavity decay rate $2\pi/(|t_1|^2+|t_2|^2+S_1)$ (agrees with our measured value prior to incorporating the membrane). (e) Dependence of resonant reflection on membrane position for the middle TEM$_{00}$ resonance in (b). The solid line shows the fit result as in (d).}
	\label{fig2}
\end{figure}

The flexural MATE system under study is drawn in Fig.~\ref{fig2}(a). The left sub-cavity comprises a flat input mirror (M1) with four 0.2-mm-diameter, 21-$\upmu$m-thick photoresist (AZ 40XT photoresist, spin coated at 3000 rpm, baked on a hot plate at 120\degC~for 240 s, exposed at 450 mJ/cm$^2$ UV, baked on a hot plate at 120\degC~for 120 s, and developed in AZ 300 MIF for 180 s) spacers positioned a distance 11 mm from the mirror center. A 24 mm $\times$ 24 mm, 0.65-mm-thick silicon chip supporting a 1 mm $\times$ 1 mm \SiN~membrane of thickness 88$\pm 3$ nm is pressed against the spacers by three piezo-driven $\sim$mm-diameter hemispherical ``pushers'' positioned a radius 3 mm from the mirror center. We choose these ``thick'' photoresist spacers to later test the durability of the chip under extreme flexure (Sec.~\ref{sec:wavelength-scale}), and to reduce sensitivity to dust particles that might preclude compression to small membrane-mirror separations. The right sub-cavity boundary is defined by a high-reflectivity backstop mirror (M2) a distance $L=10$~cm from M1, that can be swept with a second set of piezo actuators. 

We characterize the optical modes of the cavity system by reflecting 5 mW of laser light ($\lambda=1550$ nm) from M1 while sweeping M2. On resonance, the cavity's internal losses (including M2's transmission) reduce the reflected power, as shown in Fig.~\ref{fig2}(a). For convenience, we have mapped the backstop mirror displacement onto laser detuning and normalized the reflected power by its off-resonance value. The present cavity is overcoupled ($|t_1|^2$ dominates the cavity loss), verified with heterodyne analysis, and (more simply) by noting the initial reduction in resonantly reflected power when moving an absorber into the cavity mode. Figure \ref{fig2}(b) shows the same measurement (now swept along the vertical axis) over a wide range of backstop and membrane positions. Following the method of \cite{Reinhardt2016Ultralow}, we simultaneously fit the resonant detunings of 3 modes to the expression for $\omega_\text{MATE}$ in Eq.~\ref{eq:intro-omega-mate}, including a fourth-order polynomial stretching function along both axes to compensate for piezo nonidealities (part of this fit is shown as a white dashed curve). Assuming the membrane's refractive index $n_\text{SiN}=2.0$, this provides an additional estimate of the membrane's thickness $d =81.347 \pm 0.008$~nm. Despite the qualitative differences between the MATE and MIM systems' optical mode spectra, this approach still systematically underestimates $d$, as noted in MIM systems \cite{Sankey2010Strong,Reinhardt2016Ultralow}. 

The locations, depths, and widths of the observed resonances can be quantitatively captured with our 1D model, the results of which are shown in Fig.~\ref{fig2}(c). We model M1's internal loss (absorption and scattering) with a single-pass power attenuation factor $e^{-S_1}$ (where $S_1$ is a real-valued constant), at the mirror-air interface inside the cavity \cite{Stambaugh2015From}, while M2's internal loss is combined with its transmission into a larger effective $|t_2|$ (for reflection measurements, it is not useful to distinguish the two losses). We also include a ``mode-matching'' parameter $\epsilon$ defined as the fractional input power that actually couples to the TEM$_{00}$ cavity modes \cite{Hood2001Characterization}; the other fraction ($1-\epsilon$) simply reflects from the cavity while resonant with the TEM$_{00}$ modes, but couples to higher-order transverse modes at other detunings (additional faint resonances in Fig.~\ref{fig2}(b)). We assume the membrane has negligible optical losses \cite{Wilson2009Cavity,Sankey2010Strong,Reinhardt2016Ultralow}, a refractive index $n_\text{SiN}=2.0$, and thickness $d =88$~nm as measured with a white-light interferometer. Finally, by simultaneously fitting the observed dependence of linewidth (Fig.~\ref{fig2}(d)) and resonant reflection (Fig.~\ref{fig2}(e)) on membrane position, we estimate $\epsilon = 0.75\pm0.05$,  $|t_1|^2 = 7.5\pm0.3 \times 10^{-3}$, $S_1= 8.0\pm 0.8 \times 10^{-4}$, and M2 total loss $|t_2|^2 = 6\pm1 \times 10^{-4}$.  Our estimated value of $|t_1|^2$ is consistent with M1's specified power reflectivity ($>0.99$), and we infer that this simple photolithography protocol is compatible with cavity finesse of at least $2\pi/S_1 \sim 8000$. For higher-finesse systems, we recommend harder spacers fabricated with a less invasive method, such as deposition through a shadow mask.

In addition to validating the model, Fig.~\ref{fig2} illustrates that it is straightforward to tune the dissipative and dispersive optomechanical coupling over the full range by flexure, which is perhaps not surprising: in the chosen geometry, the required membrane displacement of 775 nm can be achieved with 80 m ROC, which is comparable to the natural curvature ($\sim$100 m) induced by coating one side of the Si wafer with 100 nm of stoichiometric \SiN. We also emphasize that the cavity mode frequency, decay rate, and reflected power are all observed to be smooth functions of membrane position; this is a significant departure from MIM systems, wherein avoided crossings with higher-order transverse modes tend to cause a multitude of sudden changes or discontinuities in these quantities as the membrane is swept over a similar range, even in well-aligned systems \cite{Sankey2010Nonlinear, Karuza2011Tunable, Reinhardt2016Ultralow}. Furthermore, our simple 1D model, which uses a single value of $\epsilon$ to describe the entire data set, quantitatively captures the observed (smooth) oscillations in Figs.~\ref{fig2}(d)-(e), highlighting that MATE resonances are well-approximated by a single transverse mode.

\section{Large flexure}\label{sec:wavelength-scale}

\begin{figure}[!htb]
	\centering
	\includegraphics[width=10cm]{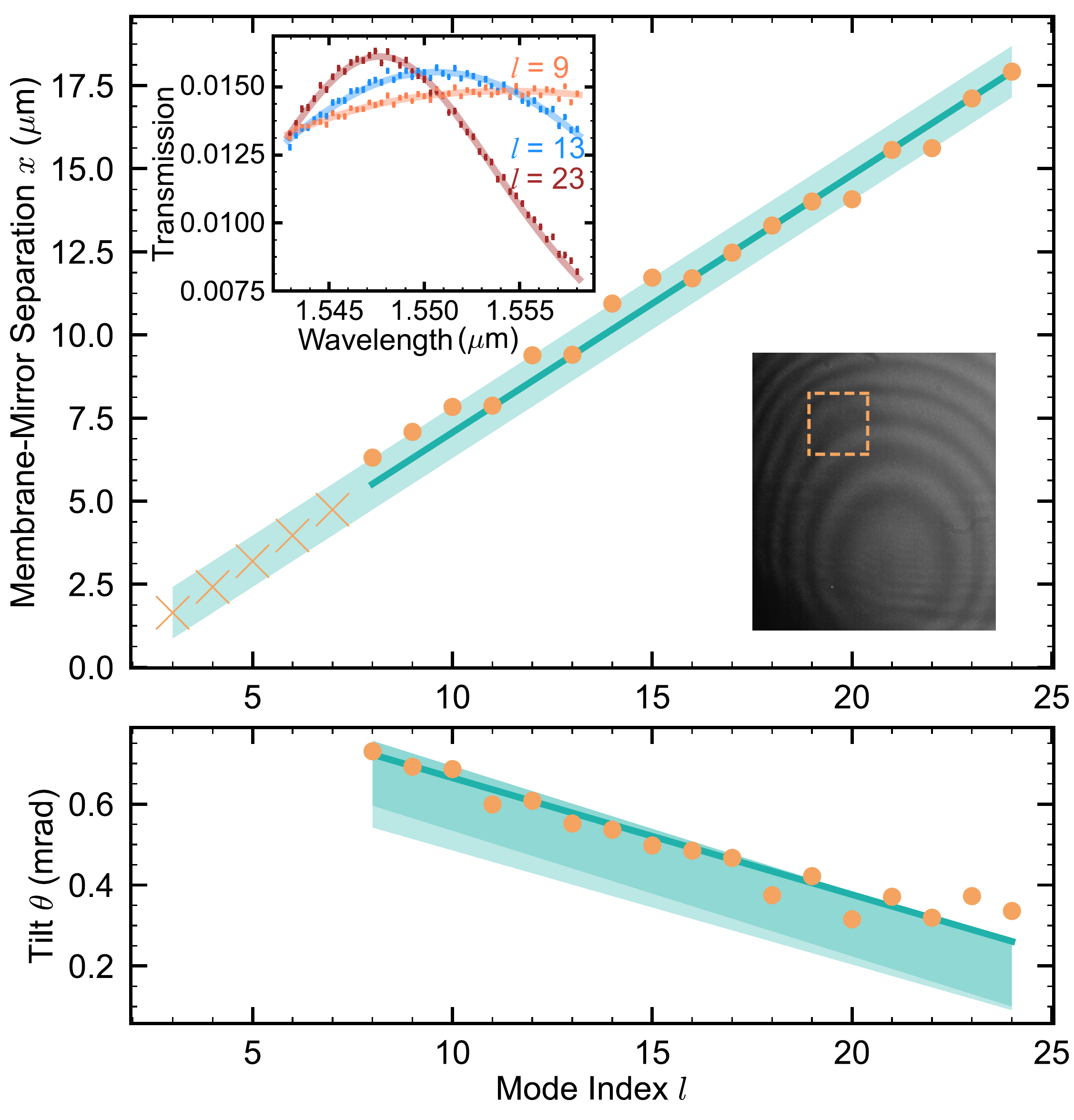}
	\caption{Large flexure. Length $x$ (top) and tilt $\theta$ (bottom) as a function of the mode index $l$. The starting index $l_0$ is estimated from a fit to the transmission spectra (upper inset) described in the main text. The uncertainty on $l_0$ is $\pm 1$, as represented by the shaded area. Below $l$=$8$, the transmission data is too broad to constrain $x$, and ``$\times$'' symbols represent the passing of a bright fringe for further flexure. The lower inset shows the interference pattern (1310 nm light) of the mirror-membrane system while maximally flexed; dotted square represents the approximate location and size of the membrane. The dark shaded area corresponds to the uncertainty in membrane thickness and M1's coating (included via transfer matrix calculation). The tilt was calculated assuming the measured beam radius $\sigma = 100$~$\mu$m, and the light shading corresponds to a conservative upper bound ($\sigma=110$~$\um$) on the actual value at the membrane.}
	\label{fig3}
\end{figure}

The chip can also be flexed far enough to contact M1, and, by engaging a single pusher, make modest adjustments to the tilt while protecting the membrane from collision. Figure \ref{fig3} summarizes our observations over the full range of displacements. To measure the membrane-mirror separation and tilt, we move the photodiode inside the cavity to measure the transmission through the stack. As discussed in Appendix~\ref{app:Pt}, the transmitted power $P_t$ depends on the membrane-mirror distance $x_0$ (at center of the incident beam) and tilt $\theta$ as 
\begin{align} 
&P_t \approx\frac{|t_1|^2|t_m|^2}{1+|r_1|^2|r_m|^2-2|r_1||r_m|\cos(2kx_0 + \phi)}\bigg\{ 1~- \nonumber\\
&~~ k^2\theta^2\sigma^2|r_1||r_m| \frac{(1+|r_1|^2|r_m|^2)\cos(2kx_0 + \phi)+|r_1||r_m|(\cos(4kx_0 + 2\phi)-3)}{[1+|r_1|^2|r_m|^2-2|r_1||r_m|\cos(2kx_0+ \phi)]^2}\bigg\} \label{eq:Pt_angle},
\end{align} 
where $k=2\pi/\lambda$ is the laser wavenumber, $\sigma=100~\um$ is the measured beam radius, and $\phi = \phi_1 + \phi_\text{m}$ is the sum of reflection coefficient phases for M1 ($\phi_1$) and the membrane ($\phi_\text{m}$). For fixed $\lambda=1550$ nm and swept $x_0$, we observe transmission peaks every $\lambda/2$, allowing simple measurement of \emph{relative} displacements, but the longitudinal index $l_0$ of the \emph{first} observed peak is not known initially. Centering $x_0$ at one such peak, we can roughly estimate $x_0$ and $\theta$ by sweeping $\lambda$ over a range that is small compared to M1's bandwidth, producing spectra like those inset to Fig.~\ref{fig3}; qualitatively, reducing $x_0$ widens the peaks while increasing $\theta$ reduces the height, as per Eq.~\ref{eq:Pt_angle}. To more accurately estimate $l_0$, we simultaneously fit \emph{all} transmission spectra as follows. We fix the mirror M1's phase to $\phi_1 = \pi$ (varying this within reason does not significantly affect our estimate), the membrane thickness $d=88$ nm, and \SiN~index $n_\text{SiN}=2$. We assume $\theta \approx \theta_0 - A \Delta x_0$ for a geometrical constant $A$ and initial tilt $\theta_0$ (at mode $l_0$), and treat the global constants $|r_1|^2$, $\theta_0$, and $A$, along with each spectrum's peak wavelength as free parameters. We then select the (integer) value of $l_0$ that minimizes $\chi^2$. Doing so yields $l_0 = 24\pm 1$, $|r_1|^2 = 0.9935 \pm 0.0001$, $\theta_0 = 0.18 \pm 0.08$ mrad, and $A = 0.040 \pm 0.006$ mrad/$\mu$m, corresponding to the solid curves in Fig.~\ref{fig3}. Uncertainties are dominated by a conservative assumed range of possible membrane thickness $d=88\pm 3$ (combined range from our white light interferometer and ellipsometer), which defines the shaded region in Fig.~\ref{fig3}(top). For tilt, the dominant uncertainties arise from both the the membrane thickness and the exact details of M1's coating (the dark shaded region includes the same fit with leading wavelength-dependence of $r_1$). The $\sim$10\% uncertainty in the beam spot-size corresponds to a (smaller) uncertainty as is shown by the light shaded area. As a consistency check, the orange points show the results of \emph{individual} transmission spectra fits assuming $|r_1|^2 =0.9935$ with $\theta$ and $l_0$ as free parameters; this also gives a sense of each spectrum's ability to resolve these quantities.

Below $l=8$, the spectra are too broad to reliably estimate the cavity length, but we count 5 more fringes (``$\times$'' symbols) before the chip contacts M1. At our furthest flexure ($l=3$), we estimate $x_0=1.64 \pm 0.78~\um$. The tilt is observed to vary linearly by 0.5 milliradians over 12.4 $\um$ (16 modes) displacement, as expected and consistent with the 3 mm lateral offset between the membrane and engaged pusher. Extrapolating to $l=3$ implies a total added tilt of $\sim$0.7 milliradians, and the maximally displaced geometry corresponds to radius of curvature $R_\text{frame}$ = 3 m.

These estimates are corroborated by shining a widely-collimated, low-coherence $\lambda=1310$ nm laser on the input mirror, and collecting the reflected interference image with a CCD, an image of which is shown in Fig.~\ref{fig3}(inset); this image is taken with the chip contacting M1, and the approximate size and location of the membrane is indicated by a dashed box. The membrane is $\sim$3 fringes from the center, suggesting a mirror-membrane separation of $\sim$2 $\um$, a few-mm offset from the point of minimal separation, and a tilt $\sim 1$ milliradians, consistent with the above estimates.

\section{Conclusions}\label{sec:conclusions}

We have demonstrated a passively aligned, flexure-tuned membrane-at-the-edge optomechanical system. Gentle flexure (80 m ROC) of the silicon chip accesses the full range of available optomechanical couplings, and extreme flexure (3 m ROC) accesses a wavelength-scale separation between a membrane and flat mirror. This monolithic, flexural approach defines a rigid mirror-membrane gap while still allowing for \emph{in situ} adjustment. By applying pressure asymmetrically, we protect the membrane from contacting the mirror while keeping the tilt to within a milliradian (even over the large membrane travel), which is sufficient to avoid significantly reducing the optomechanical coupling \cite{Sankey2010Strong}. On the other hand, for the chosen 3 mm lateral offset between the membrane and point of closest approach, we can use flexure to tune the tilt by as much as 0.7 mrad. If applied to a MIM system, this enables one to make fine \emph{in situ} adjustments to the membrane-mediated coupling between transverse cavity modes.

The relationship between the membrane's mechanical quality factor and the frame's ROC (in a vacuum system) remains to be seen. However, we expect flexure will not play an important role for gentle distortions of Fig.~\ref{fig2}, and, for high-$Q$ trampolines \cite{Reinhardt2016Ultralow,Norte2016Mechanical}, we expect the dominant effect of flexure to be increased tension in the tethers, since chip distortions will essentially serve to separate the point-like clamps rather than redistributing stress along the edge of a membrane. Note the strain associated with our most aggressive flexure is only $10^{-4}$, which would increase the \SiN stress by tens of MPa, thereby changing the resonant frequency at the percent scale.

Within the context of MATE systems, these results are of interest for generating low-noise, purely quadratic dispersive coupling, and stronger forms of quadratic dispersive or linear dissipative couplings, while (at the same time) suppressing abrupt changes in the cavity mode as the membrane moves. Equally interestingly, it should be straightforward to realize a rigid, tunable, MIM system at wavelength scale by positioning a small back mirror (e.g., a fiber mirror \cite{Hunger2010A,Flowers2012Fiber}) within the silicon etch pit. Finally, we suggest that, by instead depositing spacers on the supporting frame of a second membrane, it should be straightforward to create wavelength-scale two-membrane cavities useful for (among other things) increasing the coupling in nested cavity geometries.

\section*{Appendices}
\markright{Appendices}
\addcontentsline{toc}{section}{Appendices}
\renewcommand{\thesubsection}{\Alph{subsection}}

\subsection{Membrane in cavity: scattering model in 1D}\label{app:scattering}

In this section, we present a simple 1D scattering model describing the membrane-cavity system shown in Fig.~1(a). Section~\ref{app:scattering-setup} reviews the high-finesse (closed-cavity) approximation, Sec.~\ref{app:scattering-resonances} applies this to relate cavity length, membrane position, and the wavenumber (frequency) of resonant light, providing analytical expressions in the MIM and MATE limits. Incorporating a small end mirror transmission, we then derive expressions for the cavity's dissipation in Sec.~\ref{app:scattering-dissipation}. Finally, Secs.~\ref{app:scattering-dissipative-optomechanics} and \ref{app:scattering-dispersive-optomechanics} derive expressions for the dissipative and dispersive optomechanical couplings, discussing the advantages of the MATE system. Of note, the linear dissipative coupling can be suppressed by many orders of magnitude at locations of purely quadratic dispersive coupling, providing a potentially simpler pathway toward QND readout of mechanical energy \cite{Thompson2008Strong,Jayich2008Dispersive} in a single-port cavity, but \emph{without} the requirement \cite{Yanay2016Quantum} of strong single-photon optomechanical coupling. Additionally, for low-transmission ($t_\text{m}$) membranes (e.g., Refs.~\cite{Stambaugh2015From, Norte2016Mechanical, Chen2017High}), the single-photon strong quadratic dispersive and linear dissipative coupling parameters can both be enhanced by a factor $\sim 1/|t_\text{m}|$.

\subsubsection{Setup: high-finesse cavity-membrane system}\label{app:scattering-setup}

We consider the Fabry-Perot geometry drawn in Fig.~1(a), comprising two end mirrors (gray) separated by a distance $L$, and a partially reflective membrane (blue) positioned a distance $x$ from the first mirror. The right (left) moving free-space field amplitudes $A_j$ ($B_j$) at the membrane surfaces are related by the membrane's coefficients of transmission ($t_\text{m}$) and reflection ($r_\text{m}$) as
\begin{align}
\label{eq:A2}A_2&=t_\text{m} A_1 + r_\text{m} B_2\\
\label{eq:B1}B_1&=t_\text{m} B_2 + r_\text{m} A_1.
\end{align}
For the case of a high-finesse cavity, the end mirror reflectivities $r\approx-1$, meaning the fields are also related by the round trips to the end mirrors and back as
\begin{align}
\label{eq:A1}A_1&=-B_1e^{2ikx}\\
\label{eq:B2}B_2&=-A_2e^{2ik(L-x)},
\end{align} 
where $k=2\pi/\lambda$ is the light's wavenumber, and $\lambda$ is its wavelength. This closed-cavity (high-finesse) approximation enables us to derive the cavity's resonant frequency, dissipation, and corresponding optomechanical couplings.

\subsubsection{Resonant conditions}\label{app:scattering-resonances}

Eliminating the fields $A_j$ and $B_j$ from the above four equations yields a transcendental equation constraining the cavity length, membrane position, and wavelength:
\begin{align}
(t_\text{m}^2-r_\text{m}^2)e^{ikL}-e^{-ikL}&=2r_\text{m} \cos(2kx-kL)\label{eq:first_res}.
\end{align}
This is ``readily'' solved numerically for a membrane of arbitrary (complex-valued) reflection and transmission coefficients. However, if we assume a \emph{lossless} membrane ($|t_\text{m}|^2+|r_\text{m}|^2=1$, which is a good approximation for \SiN~at infrared wavelengths \cite{Wilson2009Cavity,Sankey2010Strong}), unitarity then imposes that $|A_1|^2+|B_2|^2=|A_2|^2+|B_1|^2$. Expressing $t_\text{m}=|t_\text{m}| e^{i\phi_\text{t}}$ and $r_\text{m}=|r_\text{m}| e^{i\phi_\text{r}}$ (for real-valued phases $\phi_\text{t}$ and $\phi_\text{r}$), this means $e^{2i(\phi_\text{t}-\phi_\text{r})}=-1$, and the above transcendental simplifies to
\begin{align}\label{eq:transcendental}
-\cos(kL+\phi_\text{r})&=|r_\text{m}|\cos(2kx-kL).
\end{align}
Expanding the cosines enables one to solve for the resonant length
\begin{align}
L&=\frac{1}{k}\arctan\left[\frac{\cos(\phi_\text{r})+|r_\text{m}|\cos(2kx)}{\sin(\phi_\text{r})-|r_\text{m}|\sin(2kx)}\right].\label{eq:res_length}
\end{align}

To derive analytical expressions for the resonant \textit{frequency} (i.e., the convenient quantity for calculating dispersive coupling) requires further approximation. We now consider the ``long cavity'' limit $L \gg \lambda/2$, such that the membrane perturbs the resonant wavenumber $k$ from that of the $N^\text{th}$ empty cavity resonance $k_N=2\pi /\lambda_N=N\pi/L$ (where $N\gg 1$, and $\lambda_N$ is the $N^\text{th}$ empty cavity resonance wavelength) by a comparatively small amount $\Delta k \ll k_N$ as
\begin{equation}\label{eq:k}
k=k_N+\Delta k.
\end{equation} 
In this limit, it is possible to derive analytical expressions for the resonant $k$ (and frequency $\omega=ck$) of MIM and MATE systems.

\paragraph{MIM:} 

Summarizing the method of \cite{Jayich2008Dispersive}, we approximate Eq.~\ref{eq:transcendental} for small displacements $\Delta x \ll L$ from the cavity center $L/2$. To leading order,
\begin{equation}
k \Delta x \approx k_N \Delta x + \Delta k \Delta x \approx k_N \Delta x,
\end{equation}
such that Eq.~\ref{eq:k} in Eq.~\ref{eq:transcendental} yields a resonant wavenumber
\begin{align}
k_\text{MIM}\approx  \frac{N\pi}{L}+\frac{1}{L}\left(\arccos\left[(-1)^{N+1}|r_\text{m}|\cos(2 k_N\Delta x)\right]- \phi_\text{r}\right).\label{eq:k_MiM}
\end{align}
and frequency 
\begin{align}\label{eq:omega_mim}
\omega_\text{MIM} \approx& N\omega_{\mathrm{FSR}} +\frac{\omega_{\mathrm{FSR}}}{\pi} \left(\arccos\left[(-1)^{N+1}|r_\text{m}|\cos(2 k_N\Delta x)\right]- \phi_\text{r}\right),
\end{align} 
where $\omega_{\mathrm{FSR}}=\pi c/L$ is the empty cavity's free spectral range (FSR). Figure 1(b) shows the $x$-dependence of $\omega_\text{MIM}$ for several values of $N$ and $r_\text{m}$. Note for these plots we have followed the convention of \cite{Thompson2008Strong}, choosing $r_\text{m}=-|r_\text{m}|$ and $t_\text{m} =i|t_\text{m}|$, to facilitate the visualization of avoided crossings. The choice of phase $\phi_\text{r}$ and $\phi_\text{t}$ shifts all modes by at most a free spectral range, and does not affect our conclusions. The convention of \cite{Jayich2008Dispersive}, e.g., is to assume an infinitesimally thin membrane of varied index. This represents a closer approximation to the perturbation from a thin dielectric slab \cite{Sankey2008Improved, Sankey2010Strong}, but for other structures, such as photonic crystal reflectors \cite{Stambaugh2015From,Chen2017High}, the phases $\phi_\text{r}$ and $\phi_\text{t}$ can vary significantly from either approximation.


\paragraph{MATE:}

If the membrane is instead positioned a distance $\Delta x\ll L$ from the left (input) mirror,
\begin{equation}\label{eq:mate-kx-approximation}
k\Delta x=k_N \Delta x+\Delta k \Delta x\approx k_N \Delta x
\end{equation} 
such that Eq.~\ref{eq:k} in Eq.~\ref{eq:res_length} yields the wavenumber
\begin{align}
k_\text{MATE}\approx \frac{N\pi}{L} + \frac{1}{L}\arctan\left[\frac{\cos(\phi_\text{r})+|r_\text{m}|\cos(2k_N \Delta x)}{\sin(\phi_\text{r})-|r_\text{m}|\sin(2k_N \Delta x)}\right]\label{k_MCTm},
\end{align} 
and resonant frequency
\begin{align}\label{eq:omega_mate}
\omega_\text{MATE} &\approx N\omega_{\mathrm{FSR}}+\frac{\omega_{\mathrm{FSR}}}{\pi}\arctan\left[\frac{\cos(\phi_\text{r})+|r_\text{m}|\cos(2k_N \Delta x)}{\sin(\phi_\text{r})-|r_\text{m}|\sin(2k_N \Delta x)}\right]. 
\end{align}
This is plotted for several values of $r_\text{m}$ and $N$ in Fig.~1(c). 

\paragraph{Comparison:}
The spectra of MIM and MATE resonances comprise a network of transmission-mediated avoided crossings between the left sub-cavity mode frequencies (having negative slope) and right sub-cavity mode frequencies (having positive slope). For the MATE case, the left sub-cavity is comparatively infinitesimal, leading to large dispersive optomechanical coupling, especially in the low-$t_\text{m}$ limit. As discussed in Sec.~\ref{app:scattering-dispersive-optomechanics}, the coupling strength is practically limited by $t_\text{m}$, which prevents light from remaining in the left sub-cavity indefinitely.

In order to derive expressions for the \emph{strong} dispersive optomechanical coupling parameters, we must first calculate the dissipation of these cavity modes.

\subsubsection{Cavity dissipation}\label{app:scattering-dissipation}

The quantities $|A_j|^2$ and $|B_j|^2$ calculated from Eqs.~\ref{eq:A2}-\ref{eq:B2} are proportional to the power landing on the end mirrors, which, combined with the left (right) end mirror transmissions $|t_1|^2$ ($|t_2|^2$), enables an estimate of the cavity's energy decay rate $\kappa$ as follows. Power circulating in the left (right) sub-cavity $P_1$ ($P_2$) corresponds to stored energy
\begin{align}\label{eq:E_total}
E&=\frac{2x}{c}P_1 + \frac{2(L-x)}{c}P_2,
\end{align}
which will leak out of the end mirrors at a rate
\begin{align}\label{eq:dtE-basic}
\partial_t E = -P_1 |t_1|^2 -P_2 |t_2|^2.
\end{align} 
We can eliminate $P_2$ from these equations by plugging Eqs.~\ref{eq:A1}-\ref{eq:B2} into Eqs.~\ref{eq:A2}-\ref{eq:B1} and taking the ratio
\begin{align}\label{eq:ratio}
\left|\frac{A_2}{B_1}\right|^2 = \frac{P_2}{P_1} =  \frac{1+|r_\text{m}|^2+2|r_\text{m}|\cos(2kx +\phi_\text{r})}{1-|r_\text{m}|^2}.
\end{align}
Substituting this into the energy (Eq.~\ref{eq:E_total}) and rearranging,
\begin{align}\label{eq:P1}
P_1 =\frac{c E}{2}\left(x + (L-x)\frac{1+|r_\text{m}|^2+2|r_\text{m}|\cos(2kx +\phi_\text{r})}{1-|r_\text{m}|^2}\right)^{-1},
\end{align}
which yields a leak rate (Eq.~\ref{eq:dtE-basic})
\begin{align}
\partial_t E= -\kappa E,
\end{align}
with
\begin{align}\label{eq:kappa(x)}
\kappa = \frac{(1-|r_\text{m}|^{2})c|t_1|^2+(1+2|r_\text{m}|\cos(2kx+\phi_\text{r})+|r_\text{m}|^{2})c|t_2|^2} {2x(1-|r_\text{m}|^{2})+2(L-x)(1+2|r_\text{m}|\cos(2kx+\phi_\text{r})+|r_\text{m}|^{2})}.
\end{align}
Note this expression is valid for any position $x$, and can be used to normalize the dissipative (Sec.~\ref{app:scattering-dissipative-optomechanics}) and dispersive (Sec.~\ref{app:scattering-dispersive-optomechanics}) couplings for either geometry.

\subsubsection{Dissipative optomechanical coupling}\label{app:scattering-dissipative-optomechanics}
Here we use Eq.~\ref{eq:kappa(x)} to extract analytical expressions for the linear dissipative strong coupling parameter \cite{Elste2009Quantum,Weiss2013Strong,Weiss2013Quantum}
\begin{align}\label{eq:Btilde-general}
\tilde{B} &\equiv \frac{1}{\kappa}\frac{d\kappa}{dx} x_\text{zpf},
\end{align}
where $x_\text{zpf}= \sqrt{\hbar/(2 m\Omega)}$ is the membrane's zero-point motion, $m$ is its mass and $\Omega$ is its resonant frequency. Reminding ourselves of the resonant wavenumber's position dependence $k(x)$, the general analytical expressions for $\tilde{B}$ are cumbersome. However, for a single-port cavity ($|t_1|^2 \approx 0$ or $|t_2|^2 \approx 0$, required, e.g., for dissipative ground state cooling \cite{Elste2009Quantum}) and a high-reflectivity membrane ($|t_\text{m}|\ll 1$), the expressions simplify dramatically. 
\paragraph{MIM:}
If $t_2=0$, the dissipative coupling (Eqs.~\ref{eq:Btilde-general} and \ref{eq:kappa(x)}) becomes
\begin{align}\label{eq:Btilde-MIM}
\tilde{B}_\text{MIM} = \frac{2|r_\text{m}|}{L} \frac{ |r_\text{m}| + \cos(2k_\text{MIM} x + \phi_\text{r}) + \left(  k_\text{MIM}+ x\frac{dk_\text{MIM}}{dx}\right)L\sin(2k_\text{MIM}x+\phi_\text{r})}{1+|r_\text{m}|\cos(2k_\text{MIM} x+\phi_\text{r})}  x_\text{zpf}.
\end{align}
In the limit $|t_\text{m}|\ll 1$ with $k_\text{MIM}$ from Eq.~\ref{eq:k_MiM}, this takes on maximal values
\begin{align}
\tilde{B}_{\text{MIM,max}} \xrightarrow[|t_\text{m}|\ll 1]{} \frac{3\sqrt{3}}{2}k_N  x_{\text{zpf}}\frac{1}{|t_\text{m}|},
\end{align}
occurring at positions 
\begin{align}
\Delta x^{(\tilde{B})}_\text{MIM,max} \xrightarrow[|t_\text{m}|\ll 1]{}  \frac{1}{2k_N}\left(j\pi+ (-1)^{j+N+1}\frac{|t_\text{m}|}{\sqrt{3}}\right)
\end{align}
that are sightly shifted slightly away from $\Delta x = j\lambda/4$ to favor the right-cavity (positive $\partial_x\omega$) mode, where $\kappa$ is somewhat lower but $\partial_x \kappa$ has not yet decreased by much.

\paragraph{MATE:}
If the membrane is instead positioned near the input mirror (keeping $t_2=0$), the approximation of Eq.~\ref{eq:mate-kx-approximation} reduces the dissipative coupling (Eq.~\ref{eq:Btilde-general}) to
\begin{align}
\tilde{B}_\text{input}\approx  \frac{2|r_\text{m}|}{L}\frac{|r_\text{m}| + \cos(2k_N \Delta x+\phi_\text{r}) +2 k_N L\sin(2k_N\Delta x+\phi_\text{r})}{1+|r_\text{m}|^2+2|r_\text{m}|\cos(2k_N\Delta x+\phi_\text{r})}x_{\mathrm{zpf}}.
\end{align}
For a high-reflectivity membrane, this takes on a maximal value 
\begin{align}
\tilde{B}_\text{MATE,max} \xrightarrow[|t_\text{m}|\ll 1]{} 4k_Nx_\text{zpf} \frac{1}{|t_\text{m}|^2}
\end{align}
at locations
\begin{align}
\Delta x^{(\tilde{B})}_\text{MATE,max} \xrightarrow[|t_\text{m}|\ll 1]{}  \frac{1}{4k_N}\left( 2\pi(2j+1) -2\phi_\text{r} \pm |t_\text{m}|^2 \right).
\end{align}

\paragraph{Comparison:}
Importantly, the MATE geometry enables a dissipative coupling enhancement as high as
\begin{align}
\frac{\tilde{B}_{\text{MATE, max}}}{\tilde{B}_{\text{MIM, max}}} \xrightarrow[|t_\text{m}|\ll 1]{} \frac{8}{3\sqrt{3}} \frac{1}{|t_\text{m}|}.
\end{align}
We emphasize that this gain is for the \emph{strong} dissipative optomechanical coupling parameter, and note that these membrane positions also have non-zero linear dispersive coupling $\partial_x\omega$ in general. Positions at which $\partial_x\omega$ vanishes are discussed at the end of the next section.

\subsubsection{Dispersive optomechanical coupling}\label{app:scattering-dispersive-optomechanics}

Here we present expressions for the linear and quadratic dispersive couplings, along with their associated strong coupling parameters. 

\paragraph{MIM:}

Equation \ref{eq:omega_mim} leads to a linear ($G^{(1)}_\text{MIM}$) and quadratic ($G^{(2)}_\text{MIM}$) dispersive couplings
\begin{align}
G^{(1)}_\text{MIM} &\equiv \partial_x\omega_\text{MIM}=(-1)^{N+1}\frac{2\omega_{\mathrm{FSR}}k_N}{\pi}\frac{|r_\text{m}|\sin(2k_N \Delta x)}{\sqrt{1-|r_\text{m}|^2\cos^2(2k_N\Delta x)}}\\
G^{(2)}_\text{MIM} &\equiv \partial_x^2\omega_\text{MIM}=(-1)^{N+1}\frac{4\omega_{\mathrm{FSR}}k_N^2}{\pi}\frac{|r_\text{m}|(1-|r_\text{m}|^2)\cos(2k_N\Delta x)}{(1-|r_\text{m}|^2\cos^2(2k_N\Delta x))^{3/2}},
\end{align} 
with extremal values
\begin{align}
G^{(1)}_\text{MIM,max} &= \pm \frac{2ck_N}{L}|r_\text{m}|\\
G^{(2)}_\text{MIM,max} &= \pm \frac{4ck_N^2}{L}\frac{|r_\text{m}|}{\sqrt{1-|r_\text{m}|^2}} \xrightarrow[|t_\text{m}|\ll 1]{}  \pm \frac{4ck_N^2}{L} \frac{1}{|t_\text{m}|} \label{eq:G2,MIMmax}
\end{align}
at locations
\begin{align}
\Delta x^{(1)}_\text{MIM,max} &=  (2j+1) \frac{\lambda}{8} \\
\Delta x^{(2)}_\text{MIM,max} &= j\frac{\lambda}{4},\label{eq:MIM-xquadratic}
\end{align}
respectively ($j\in\mathbb{Z}$). For this geometry, the maximal values for quadratic coupling occur when the membrane is at a node or antinode of the empty cavity field, and the maximal linear coupling occurs at the midpoints in between. Also, the maximal values for each occur when the coupling of the other is zero, as can be seen in Fig.~1(b).

\paragraph{MATE:}

Equation \ref{eq:omega_mate} leads to linear ($G^{(1)}_\text{MATE}$) and quadratic ($G^{(2)}_\text{MATE}$) dispersive couplings
\begin{align}
G^{(1)}_\text{MATE} &\equiv \partial_x \omega_\text{MATE}=\frac{2k_N }{\pi} \omega_{\mathrm{FSR}}   \frac{|r_\text{m}|(|r_\text{m}|+\cos(2k_N\Delta x+\phi_\text{r}))}{|r_\text{m}|^2+2|r_\text{m}|\cos(2k_N\Delta x+\phi_\text{r})+1}\\
G^{(2)}_\text{MATE} &\equiv \partial^2_x \omega_\text{MATE} =-\frac{4k_N^2}{\pi}\omega_{\mathrm{FSR}}\frac{|r_\text{m}|(1-|r_\text{m}|^2)\sin(2k_N \Delta x+\phi_\text{r})}{(|r_\text{m}|^2+2|r_\text{m}|\cos(2k_N \Delta x+\phi_\text{r})+1)^2},
\end{align}
with extremal values
\begin{align}
G^{(1)}_\text{MATE,max} &=  - \frac{2c k_N}{L} \frac{|r_\text{m}|}{1-|r_\text{m}|} \xrightarrow[|t_\text{m}|\ll 1]{}  - \frac{4c k_N}{L} \frac{1}{|t_\text{m}|^2} \\
G^{(2)}_\text{MATE,max} &= G^{(2)}_\text{MATE} (\Delta x^{(2)}_\text{MATE,max}) \xrightarrow[|t_\text{m}|\ll 1]{}  \pm \frac{18}{\sqrt{3}} \frac{ck_N^2}{L} \frac{1}{|t_\text{m}|^4}
\end{align}
at locations
\begin{align}
\Delta x^{(1)}_\text{MATE,max} &=  \frac{(2j +1)\pi-\phi_\text{r}}{2k_N}  \\
\Delta x^{(2)}_\text{MATE,max} &= \frac{1}{2k_N} \Bigg (  2\pi j - \phi_\text{r}  \pm  2\arctan\Bigg\{\sqrt{\frac{6|r_\text{m}|+\sqrt{|r_\text{m}|^4+34|r_\text{m}|^2+1}}{(1-|r_\text{m}|)^2}} \Bigg\} \Bigg)
\end{align}
respectively ($j\in\mathbb{Z}$). 
Note the maximal quadratic coupling does \text{not} occur at the extrema in $\omega_\text{MATE}$ (where $G^{(1)}_\text{MATE}=0$). At those ``purely'' quadratic locations
%
%
%
%
%
\begin{align}
\Delta x^{(2)}_\text{MATE,pure} = \frac{\pm \arccos\{-|r_\text{m}|\} -\phi_\text{r} + 2\pi j}{2k_N},\label{eq:MATE-xquadratic}
\end{align}
the coupling is identical to that of the MIM system (Eq.~\ref{eq:G2,MIMmax}). 

\paragraph{Comparison:}
As noted elsewhere \cite{Purdy2012Cavity}, the maximal linear coupling is larger than that of the MIM system by a factor
\begin{equation}\label{eq:ratio-MIM-MATE-linear}
\frac{G^{(1)}_\text{MATE,max}}{G^{(1)}_\text{MIM,max}} = \frac{1}{1-|r_\text{m}|} \xrightarrow[|t_\text{m}|\ll 1]{}  \frac{2}{|t_\text{m}|^2},	
\end{equation}
but it comes at the expense of larger $\kappa$, due to the associated cavity mode's higher intensity in the (shorter) cavity (see inset of Fig.~1(c)). As such, there is no difference in the maximum possible strong coupling parameter $\tilde{A}^{(1)} =  -G^{(1)} x_\text{zpf}/\kappa$ \cite{Purdy2012Cavity}, and
\begin{align}
\tilde{A}^{(1)}_{\text{max, MIM}} =  \tilde{A}^{(1)}_{\text{max, MATE}} \xrightarrow[|t_\text{m}|\ll 1]{}  -\frac{8 k_N x_{\mathrm{zpf}}}{|t_1|^2|t_\text{m}|^2}.
\end{align}
On the other hand, the maximal \emph{quadratic} coupling can be enhanced by a \emph{larger} factor
\begin{equation}\label{eq:ratio-MIM-MATE-quadratic}
\frac{G^{(2)}_\text{MATE,max}}{G^{(2)}_\text{MIM,max}} \xrightarrow[|t_\text{m}|\ll 1]{} \frac{9}{2\sqrt{3}} \frac{1}{|t_\text{m}|^3}.
\end{equation} 
This improvement also comes with increased $\kappa$, but the strong coupling parameter ($\tilde{A}^{(2)} = -G^{(2)} x_\text{zpf}^2 / \kappa$) can still be improved, as
\begin{align}
\frac{\tilde{A}^{(2)}_\text{max,MATE}}{\tilde{A}^{(2)}_\text{max,MIM}} \xrightarrow[|t_\text{m}|\ll 1]{}&  \frac{4}{3\sqrt{3}}\frac{1}{|t_\text{m}|}.
\end{align} 
We note the caveat that these MATE expressions assume the limit $4\Delta x/L \ll |t_\text{m}|^2$, such that the membrane's transmission limits the coupling rates (i.e., not the mirror-membrane separation).

\paragraph{Purely quadratic dispersive coupling:}

When the membrane resides at locations having purely quadratic dispersive coupling (to lowest order), the remaining position dependence of the cavity mode still represents a linear back-action that can preclude quantum nondemolition (QND) readout of mechanical energy eigenstates \cite{Miao2009Standard}, even in an ideal one-port cavity \cite{Yanay2016Quantum}. For a MIM system with $t_2=0$ (no restrictions on membrane reflectivity), Eqs.~\ref{eq:MIM-xquadratic} and \ref{eq:Btilde-MIM} yield a dissipative coupling 
\begin{align}
\tilde{B}_\text{MIM,pure} &= \pm 2k_N x_{\mathrm{zpf}} \frac{|r_\text{m}|}{|t_\text{m}|}\label{eq:B-mim-xq}
\end{align} 
at these ``purely quadratic'' points. For the MATE system with the membrane near the \emph{input} mirror, the dissipative coupling at these points is 
\begin{align}
\tilde{B}_\text{MATE,pure} &=  \pm 4k_N x_{\mathrm{zpf}} \frac{|r_\text{m}|}{|t_\text{m}|}\label{eq:B-mate-xq}
\end{align}
which is twice as large as for the MIM system. More compellingly, if the membrane is positioned near the \emph{back} mirror, where $L-x\ll L$, the dissipative coupling at these points
\begin{align}
\tilde{B}^\prime_\text{MATE,pure} = \pm 4k_Nx_{\text{zpf}}  \frac{\Delta x_{\text{MATE, pure}}}{L} \frac{|r_\text{m}|}{|t_\text{m}|},
\end{align}
a reduction by the factor $2\Delta x_{\text{MATE,pure}}/L$ relative to the MIM case discussed in \cite{Miao2009Standard, Yanay2016Quantum}, thereby suppressing the unwanted linear back-action by many orders of magnitude.

\subsection{Fabry-Perot cavity with tilted end mirror}\label{app:Pt}

Here we derive an expression for the power transmitted through a wedged Fabry-Perot cavity comprising mirror M1 and the membrane in Fig.~1(a). Section \ref{app:empty-cavity} reviews a 1D empty cavity (parallel mirrors, plane waves), and Section \ref{app:empty-cavity-tilt} treats a wedged cavity as a continuum of such 1D cavities, all having different lengths.

\subsubsection{Empty cavity in 1D}\label{app:empty-cavity}

For the mirror-membrane cavity in Fig.~1(a), we can relate the transmitted field $A_2$ to the input field $E_\text{in}$ from the left by summing all possible paths the light can take to exit through the membrane. Specifically, every path is transmitted through M1, travels the length $x$ of the cavity, and can make any number $j$ of round trips (reflecting off both mirrors each time) before transmitting through the membrane. Summing all such paths yields
\begin{align}
A_2 &= E_\text{in}(t_1t_\text{m}e^{ikx})\sum_{j=0}^\infty\left[r_1r_\text{m}e^{2ikx}\right]^j\\
&=E_\text{in}\frac{t_1t_\text{m}e^{ikx}}{1 - r_1r_\text{m}e^{2ikx}}\label{eq:Et},
\end{align} 
where $k$ is the incident light's wavenumber, and we have yet made no assumption about $r_1$, $t_1$, $r_\text{m}$ and $t_\text{m}$.

\subsubsection{Empty cavity in 2D with tilt}\label{app:empty-cavity-tilt}

In this section, we derive the transmission through a mirror-membrane stack when the two surfaces are not parallel, such that the cavity length varies with the transverse position $y$ as $x=x_0 + \theta y$, where $x_0$ is the nominal separation and $\theta$ is the slope. For convenience, consider an incident Gaussian beam having power density $\sqrt{\frac{2}{\pi\sigma^2 }}e^{-2y^2/\sigma^2}$ (i.e., normalized to total power 1), with beam radius $\sigma$. At a given transverse location $y$, Eq.~\ref{eq:Et} then yields a fractional power transmission
\begin{align}\label{eq:pt}
p_t &\equiv |A_2/E_\text{in}|^2 = \frac{|t_1|^2|t_\text{m}|^2}{1+|r_1|^2|r_\text{m}|^2-2|r_1||r_\text{m}|\cos(2kx_0+2k\theta y+\phi)}\sqrt{\frac{2}{\pi\sigma^2 }}e^{-\frac{2y^2}{\sigma^2}},
\end{align} 
where we now employ the convention $r_1 = |r_1|e^{i\phi_1}$ ($r_\text{m} = |r_\text{m}|e^{i\phi_\text{m}}$), with $\phi_1$ ($\phi_\text{m}$) being the real-valued phase of M1 (the membrane), and $\phi\equiv \phi_1+\phi_m$.

To estimate the total fractional transmission $P_t$, we can now integrate Eq.~\ref{eq:pt} over $y$. To simplify, we assume the tilt is sufficiently small that $2k\theta y\ll 1$, and expand the previous expression over the whole beam, yielding
\begin{align} 
p_t& \approx  \frac{|t_1|^2|t_\text{m}|^2}{1+|r_1|^2|r_\text{m}|^2-2|r_1||r_\text{m}|\cos(2kx_0 + \phi)}\sqrt{\frac{2}{\pi\sigma^2 }}e^{-\frac{2y^2}{\sigma^2}} \bigg\{1  - \nonumber \\   
& 4k^2\theta^2y^2|r_1||r_\text{m}|\frac{(1+|r_1|^2|r_\text{m}|^2)\cos(2kx_0 + \phi)+|r_1||r_\text{m}|(\cos(4kx_0 + 2\phi)-3)}{[1+|r_1|^2|r_\text{m}|^2-2|r_1||r_\text{m}|\cos(2kx_0 + \phi)]^2}\bigg \},
\end{align} 
where we have left out the term $\propto y$ for brevity since it will vanish by symmetry. Integrating over $y$ yields
\begin{align} 
P_t &=\frac{|t_1|^2|t_\text{m}|^2}{1+|r_1|^2|r_\text{m}|^2-2|r_1||r_\text{m}|\cos(2kx_0 + \phi)}  \bigg \{ 1 - \nonumber\\ &k^2\theta^2\sigma^2|r_1||r_\text{m}|\frac{(1+|r_1|^2|r_\text{m}|^2)\cos(2kx_0 +\phi)+|r_1||r_\text{m}|(\cos(4kx_0 + 2\phi)-3)}{[1+|r_1|^2|r_\text{m}|^2-2|r_1||r_\text{m}|\cos(2kx_0 + \phi)]^2}\bigg\}.
\end{align} 
Most relevantly, this expression describes how tilt reduces and broadens the resonant transmission peaks (occurring at positions satisfying $2kx_0 +\phi \approx 2\pi l$ for integer $l$), as shown in Fig.~3(inset).

\section*{Funding}
NSERC (RGPIN 418459-2012 \& 2018-05635); FRQNT (NC-172619); CFI (228130); CRC (235060); INTRIQ; Centre for the Physics of Materials at McGill.

\section*{Acknowledgments}
We thank Raphael Saint-Gelais, Yannik Fontana, Erika Janitz, Sasa Ristic and John Li for useful discussions regarding fabrication and optical characterization. VD acknowledges support from the NSERC CGS-M and FRQNT-B1 scholarships. VD and SB acknowledge support from FRQNT-B2 scholarships. SB and CR acknowledge support from Schulich fellowships.

\section*{Disclosures}
The authors declare that there are no conflicts of interest related to this article.

\addcontentsline{toc}{section}{References}
\bibliographystyle{bibstyle-jack}
\bibliography{errthing}

\begin{thebibliography}{52}%
\makeatletter
\providecommand \@ifxundefined [1]{%
 \@ifx{#1\undefined}
}%
\providecommand \@ifnum [1]{%
 \ifnum #1\expandafter \@firstoftwo
 \else \expandafter \@secondoftwo
 \fi
}%
\providecommand \@ifx [1]{%
 \ifx #1\expandafter \@firstoftwo
 \else \expandafter \@secondoftwo
 \fi
}%
\providecommand \natexlab [1]{#1}%
\providecommand \enquote  [1]{``#1''}%
\providecommand \bibnamefont  [1]{#1}%
\providecommand \bibfnamefont [1]{#1}%
\providecommand \citenamefont [1]{#1}%
\providecommand \href@noop [0]{\@secondoftwo}%
\providecommand \href [0]{\begingroup \@sanitize@url \@href}%
\providecommand \@href[1]{\@@startlink{#1}\@@href}%
\providecommand \@@href[1]{\endgroup#1\@@endlink}%
\providecommand \@sanitize@url [0]{\catcode `\\12\catcode `\$12\catcode
  `\&12\catcode `\#12\catcode `\^12\catcode `\_12\catcode `\%12\relax}%
\providecommand \@@startlink[1]{}%
\providecommand \@@endlink[0]{}%
\providecommand \url  [0]{\begingroup\@sanitize@url \@url }%
\providecommand \@url [1]{\endgroup\@href {#1}{\urlprefix }}%
\providecommand \urlprefix  [0]{URL }%
\providecommand \Eprint [0]{\href }%
\providecommand \doibase [0]{http://dx.doi.org/}%
\providecommand \selectlanguage [0]{\@gobble}%
\providecommand \bibinfo  [0]{\@secondoftwo}%
\providecommand \bibfield  [0]{\@secondoftwo}%
\providecommand \translation [1]{[#1]}%
\providecommand \BibitemOpen [0]{}%
\providecommand \bibitemStop [0]{}%
\providecommand \bibitemNoStop [0]{.\EOS\space}%
\providecommand \EOS [0]{\spacefactor3000\relax}%
\providecommand \BibitemShut  [1]{\csname bibitem#1\endcsname}%
\let\auto@bib@innerbib\@empty
\bibitem [{\citenamefont {Aspelmeyer}\ \emph {et~al.}(2014)\citenamefont
  {Aspelmeyer}, \citenamefont {Kippenberg},\ and\ \citenamefont
  {Marquardt}}]{Aspelmeyer2014Cavity}%
  \BibitemOpen
  \bibfield  {author} {\bibinfo {author} {\bibfnamefont {M.}~\bibnamefont
  {Aspelmeyer}}, \bibinfo {author} {\bibfnamefont {T.~J.}\ \bibnamefont
  {Kippenberg}}, \ and\ \bibinfo {author} {\bibfnamefont {F.}~\bibnamefont
  {Marquardt}},\ }\bibfield  {title} {\enquote {\bibinfo {title} {{Cavity
  optomechanics}},}\ }\href {http://dx.doi.org/10.1103/RevModPhys.86.1391}
  {\bibfield  {journal} {\bibinfo  {journal} {Rev. Modern Phys.}\ }\textbf
  {\bibinfo {volume} {86}},\ \bibinfo {pages} {1391} (\bibinfo {year}
  {2014})}\BibitemShut {NoStop}%
\bibitem [{\citenamefont {Thompson}\ \emph {et~al.}(2008)\citenamefont
  {Thompson}, \citenamefont {Zwickl}, \citenamefont {Jayich}, \citenamefont
  {Marquardt}, \citenamefont {Girvin},\ and\ \citenamefont
  {Harris}}]{Thompson2008Strong}%
  \BibitemOpen
  \bibfield  {author} {\bibinfo {author} {\bibfnamefont {J.~D.}\ \bibnamefont
  {Thompson}}, \bibinfo {author} {\bibfnamefont {B.~M.}\ \bibnamefont
  {Zwickl}}, \bibinfo {author} {\bibfnamefont {A.~M.}\ \bibnamefont {Jayich}},
  \bibinfo {author} {\bibfnamefont {F.}~\bibnamefont {Marquardt}}, \bibinfo
  {author} {\bibfnamefont {S.~M.}\ \bibnamefont {Girvin}}, \ and\ \bibinfo
  {author} {\bibfnamefont {J.~G.~E.}\ \bibnamefont {Harris}},\ }\bibfield
  {title} {\enquote {\bibinfo {title} {{Strong dispersive coupling of a
  high-finesse cavity to a micromechanical membrane}},}\ }\href
  {http://dx.doi.org/10.1038/nature06715} {\bibfield  {journal} {\bibinfo
  {journal} {Nature}\ }\textbf {\bibinfo {volume} {452}},\ \bibinfo {pages}
  {72} (\bibinfo {year} {2008})}\BibitemShut {NoStop}%
\bibitem [{\citenamefont {Bhattacharya}\ \emph {et~al.}(2008)\citenamefont
  {Bhattacharya}, \citenamefont {Uys},\ and\ \citenamefont
  {Meystre}}]{Bhattacharya2008Optomechanical}%
  \BibitemOpen
  \bibfield  {author} {\bibinfo {author} {\bibfnamefont {M.}~\bibnamefont
  {Bhattacharya}}, \bibinfo {author} {\bibfnamefont {H.}~\bibnamefont {Uys}}, \
  and\ \bibinfo {author} {\bibfnamefont {P.}~\bibnamefont {Meystre}},\
  }\bibfield  {title} {\enquote {\bibinfo {title} {{Optomechanical trapping and
  cooling of partially reflective mirrors}},}\ }\href
  {http://dx.doi.org/10.1103/PhysRevA.77.033819} {\bibfield  {journal}
  {\bibinfo  {journal} {Phys. Rev. A}\ }\textbf {\bibinfo {volume} {77}},\
  \bibinfo {pages} {033819} (\bibinfo {year} {2008})}\BibitemShut {NoStop}%
\bibitem [{\citenamefont {Ghadimi}\ \emph {et~al.}(2018)\citenamefont
  {Ghadimi}, \citenamefont {Fedorov}, \citenamefont {Engelsen}, \citenamefont
  {Bereyhi}, \citenamefont {Schilling}, \citenamefont {Wilson},\ and\
  \citenamefont {Kippenberg}}]{Ghadimi2018Elastic}%
  \BibitemOpen
  \bibfield  {author} {\bibinfo {author} {\bibfnamefont {A.~H.}\ \bibnamefont
  {Ghadimi}}, \bibinfo {author} {\bibfnamefont {S.~A.}\ \bibnamefont
  {Fedorov}}, \bibinfo {author} {\bibfnamefont {N.~J.}\ \bibnamefont
  {Engelsen}}, \bibinfo {author} {\bibfnamefont {M.~J.}\ \bibnamefont
  {Bereyhi}}, \bibinfo {author} {\bibfnamefont {R.}~\bibnamefont {Schilling}},
  \bibinfo {author} {\bibfnamefont {D.~J.}\ \bibnamefont {Wilson}}, \ and\
  \bibinfo {author} {\bibfnamefont {T.~J.}\ \bibnamefont {Kippenberg}},\
  }\bibfield  {title} {\enquote {\bibinfo {title} {{Elastic strain engineering
  for ultralow mechanical dissipation.}}}\ }\href
  {http://dx.doi.org/10.1126/science.aar6939} {\bibfield  {journal} {\bibinfo
  {journal} {Science}\ }\textbf {\bibinfo {volume} {360}},\ \bibinfo {pages}
  {764} (\bibinfo {year} {2018})}\BibitemShut {NoStop}%
\bibitem [{\citenamefont {Reinhardt}\ \emph {et~al.}(2016)\citenamefont
  {Reinhardt}, \citenamefont {M{\"{u}}ller}, \citenamefont {Bourassa},\ and\
  \citenamefont {Sankey}}]{Reinhardt2016Ultralow}%
  \BibitemOpen
  \bibfield  {author} {\bibinfo {author} {\bibfnamefont {C.}~\bibnamefont
  {Reinhardt}}, \bibinfo {author} {\bibfnamefont {T.}~\bibnamefont
  {M{\"{u}}ller}}, \bibinfo {author} {\bibfnamefont {A.}~\bibnamefont
  {Bourassa}}, \ and\ \bibinfo {author} {\bibfnamefont {J.}~\bibnamefont
  {Sankey}},\ }\bibfield  {title} {\enquote {\bibinfo {title} {{Ultralow-noise
  SiN trampoline resonators for sensing and optomechanics}},}\ }\href
  {http://dx.doi.org/10.1103/PhysRevX.6.021001} {\bibfield  {journal} {\bibinfo
   {journal} {Phys. Rev. X}\ }\textbf {\bibinfo {volume} {6}},\ \bibinfo
  {pages} {021001} (\bibinfo {year} {2016})}\BibitemShut {NoStop}%
\bibitem [{\citenamefont {Norte}\ \emph {et~al.}(2016)\citenamefont {Norte},
  \citenamefont {Moura},\ and\ \citenamefont
  {Gr{\"{o}}blacher}}]{Norte2016Mechanical}%
  \BibitemOpen
  \bibfield  {author} {\bibinfo {author} {\bibfnamefont {R.~A.}\ \bibnamefont
  {Norte}}, \bibinfo {author} {\bibfnamefont {J.~P.}\ \bibnamefont {Moura}}, \
  and\ \bibinfo {author} {\bibfnamefont {S.}~\bibnamefont {Gr{\"{o}}blacher}},\
  }\bibfield  {title} {\enquote {\bibinfo {title} {{Mechanical resonators for
  quantum optomechanics experiments at room temperature}},}\ }\href
  {http://dx.doi.org/10.1103/PhysRevLett.116.147202} {\bibfield  {journal}
  {\bibinfo  {journal} {Phys. Rev. Lett.}\ }\textbf {\bibinfo {volume} {116}},\
  \bibinfo {pages} {147202} (\bibinfo {year} {2016})}\BibitemShut {NoStop}%
\bibitem [{\citenamefont {Tsaturyan}\ \emph {et~al.}(2017)\citenamefont
  {Tsaturyan}, \citenamefont {Barg}, \citenamefont {Polzik},\ and\
  \citenamefont {Schliesser}}]{Tsaturyan2017Ultracoherent}%
  \BibitemOpen
  \bibfield  {author} {\bibinfo {author} {\bibfnamefont {Y.}~\bibnamefont
  {Tsaturyan}}, \bibinfo {author} {\bibfnamefont {A.}~\bibnamefont {Barg}},
  \bibinfo {author} {\bibfnamefont {E.~S.}\ \bibnamefont {Polzik}}, \ and\
  \bibinfo {author} {\bibfnamefont {A.}~\bibnamefont {Schliesser}},\ }\bibfield
   {title} {\enquote {\bibinfo {title} {{Ultracoherent nanomechanical
  resonators via soft clamping and dissipation dilution}},}\ }\href
  {http://dx.doi.org/10.1038/nnano.2017.101} {\bibfield  {journal} {\bibinfo
  {journal} {Nat. Nano.}\ }\textbf {\bibinfo {volume} {12}},\ \bibinfo {pages}
  {776} (\bibinfo {year} {2017})}\BibitemShut {NoStop}%
\bibitem [{\citenamefont {Wilson}\ \emph {et~al.}(2009)\citenamefont {Wilson},
  \citenamefont {Regal}, \citenamefont {Papp},\ and\ \citenamefont
  {Kimble}}]{Wilson2009Cavity}%
  \BibitemOpen
  \bibfield  {author} {\bibinfo {author} {\bibfnamefont {D.~J.}\ \bibnamefont
  {Wilson}}, \bibinfo {author} {\bibfnamefont {C.~A.}\ \bibnamefont {Regal}},
  \bibinfo {author} {\bibfnamefont {S.~B.}\ \bibnamefont {Papp}}, \ and\
  \bibinfo {author} {\bibfnamefont {H.~J.}\ \bibnamefont {Kimble}},\ }\bibfield
   {title} {\enquote {\bibinfo {title} {{Cavity optomechanics with
  stoichiometric SiN films}},}\ }\href
  {http://dx.doi.org/10.1103/physrevlett.103.207204} {\bibfield  {journal}
  {\bibinfo  {journal} {Phys. Rev. Lett.}\ }\textbf {\bibinfo {volume} {103}},\
  \bibinfo {pages} {207204} (\bibinfo {year} {2009})}\BibitemShut {NoStop}%
\bibitem [{\citenamefont {Sankey}\ \emph
  {et~al.}(2010{\natexlab{a}})\citenamefont {Sankey}, \citenamefont {Yang},
  \citenamefont {Zwickl}, \citenamefont {Jayich},\ and\ \citenamefont
  {Harris}}]{Sankey2010Strong}%
  \BibitemOpen
  \bibfield  {author} {\bibinfo {author} {\bibfnamefont {J.}~\bibnamefont
  {Sankey}}, \bibinfo {author} {\bibfnamefont {C.}~\bibnamefont {Yang}},
  \bibinfo {author} {\bibfnamefont {B.}~\bibnamefont {Zwickl}}, \bibinfo
  {author} {\bibfnamefont {A.}~\bibnamefont {Jayich}}, \ and\ \bibinfo {author}
  {\bibfnamefont {J.}~\bibnamefont {Harris}},\ }\bibfield  {title} {\enquote
  {\bibinfo {title} {{Strong and tunable nonlinear optomechanical coupling in a
  low-loss system}},}\ }\href {http://dx.doi.org/10.1038/nphys1707} {\bibfield
  {journal} {\bibinfo  {journal} {Nat. Phys.}\ }\textbf {\bibinfo {volume}
  {6}},\ \bibinfo {pages} {707} (\bibinfo {year}
  {2010}{\natexlab{a}})}\BibitemShut {NoStop}%
\bibitem [{\citenamefont {Purdy}\ \emph {et~al.}(2015)\citenamefont {Purdy},
  \citenamefont {Yu}, \citenamefont {Kampel}, \citenamefont {Peterson},
  \citenamefont {Cicak}, \citenamefont {Simmonds},\ and\ \citenamefont
  {Regal}}]{Purdy2015Optomechanical}%
  \BibitemOpen
  \bibfield  {author} {\bibinfo {author} {\bibfnamefont {T.~P.}\ \bibnamefont
  {Purdy}}, \bibinfo {author} {\bibfnamefont {P.-L.}\ \bibnamefont {Yu}},
  \bibinfo {author} {\bibfnamefont {N.~S.}\ \bibnamefont {Kampel}}, \bibinfo
  {author} {\bibfnamefont {R.~W.}\ \bibnamefont {Peterson}}, \bibinfo {author}
  {\bibfnamefont {K.}~\bibnamefont {Cicak}}, \bibinfo {author} {\bibfnamefont
  {R.~W.}\ \bibnamefont {Simmonds}}, \ and\ \bibinfo {author} {\bibfnamefont
  {C.~A.}\ \bibnamefont {Regal}},\ }\bibfield  {title} {\enquote {\bibinfo
  {title} {{Optomechanical Raman-ratio thermometry}},}\ }\href
  {http://dx.doi.org/10.1103/PhysRevA.92.031802} {\bibfield  {journal}
  {\bibinfo  {journal} {Phys. Rev. A}\ }\textbf {\bibinfo {volume} {92}},\
  \bibinfo {pages} {031802} (\bibinfo {year} {2015})}\BibitemShut {NoStop}%
\bibitem [{\citenamefont {Underwood}\ \emph {et~al.}(2015)\citenamefont
  {Underwood}, \citenamefont {Mason}, \citenamefont {Lee}, \citenamefont {Xu},
  \citenamefont {Jiang}, \citenamefont {Shkarin}, \citenamefont {B{\o}rkje},
  \citenamefont {Girvin},\ and\ \citenamefont
  {Harris}}]{Underwood2015Measurement}%
  \BibitemOpen
  \bibfield  {author} {\bibinfo {author} {\bibfnamefont {M.}~\bibnamefont
  {Underwood}}, \bibinfo {author} {\bibfnamefont {D.}~\bibnamefont {Mason}},
  \bibinfo {author} {\bibfnamefont {D.}~\bibnamefont {Lee}}, \bibinfo {author}
  {\bibfnamefont {H.}~\bibnamefont {Xu}}, \bibinfo {author} {\bibfnamefont
  {L.}~\bibnamefont {Jiang}}, \bibinfo {author} {\bibfnamefont {A.~B.}\
  \bibnamefont {Shkarin}}, \bibinfo {author} {\bibfnamefont {K.}~\bibnamefont
  {B{\o}rkje}}, \bibinfo {author} {\bibfnamefont {S.~M.}\ \bibnamefont
  {Girvin}}, \ and\ \bibinfo {author} {\bibfnamefont {J.~G.~E.}\ \bibnamefont
  {Harris}},\ }\bibfield  {title} {\enquote {\bibinfo {title} {{Measurement of
  the motional sidebands of a nanogram-scale oscillator in the quantum
  regime}},}\ }\href {http://dx.doi.org/10.1103/PhysRevA.92.061801} {\bibfield
  {journal} {\bibinfo  {journal} {Phys. Rev. A}\ }\textbf {\bibinfo {volume}
  {92}},\ \bibinfo {pages} {061801} (\bibinfo {year} {2015})}\BibitemShut
  {NoStop}%
\bibitem [{\citenamefont {Peterson}\ \emph {et~al.}(2016)\citenamefont
  {Peterson}, \citenamefont {Purdy}, \citenamefont {Kampel}, \citenamefont
  {Andrews}, \citenamefont {Yu}, \citenamefont {Lehnert},\ and\ \citenamefont
  {Regal}}]{Peterson2016Laser}%
  \BibitemOpen
  \bibfield  {author} {\bibinfo {author} {\bibfnamefont {R.~W.}\ \bibnamefont
  {Peterson}}, \bibinfo {author} {\bibfnamefont {T.~P.}\ \bibnamefont {Purdy}},
  \bibinfo {author} {\bibfnamefont {N.~S.}\ \bibnamefont {Kampel}}, \bibinfo
  {author} {\bibfnamefont {R.~W.}\ \bibnamefont {Andrews}}, \bibinfo {author}
  {\bibfnamefont {P.~L.}\ \bibnamefont {Yu}}, \bibinfo {author} {\bibfnamefont
  {K.~W.}\ \bibnamefont {Lehnert}}, \ and\ \bibinfo {author} {\bibfnamefont
  {C.~A.}\ \bibnamefont {Regal}},\ }\bibfield  {title} {\enquote {\bibinfo
  {title} {{Laser cooling of a micromechanical membrane to the quantum
  backaction limit}},}\ }\href
  {http://dx.doi.org/10.1103/PhysRevLett.116.063601} {\bibfield  {journal}
  {\bibinfo  {journal} {Phys. Rev. Lett.}\ }\textbf {\bibinfo {volume} {116}},\
  \bibinfo {pages} {063601} (\bibinfo {year} {2016})}\BibitemShut {NoStop}%
\bibitem [{\citenamefont {Purdy}\ \emph
  {et~al.}(2013{\natexlab{a}})\citenamefont {Purdy}, \citenamefont {Peterson},\
  and\ \citenamefont {Regal}}]{Purdy2013Observation}%
  \BibitemOpen
  \bibfield  {author} {\bibinfo {author} {\bibfnamefont {T.~P.}\ \bibnamefont
  {Purdy}}, \bibinfo {author} {\bibfnamefont {R.~W.}\ \bibnamefont {Peterson}},
  \ and\ \bibinfo {author} {\bibfnamefont {C.~A.}\ \bibnamefont {Regal}},\
  }\bibfield  {title} {\enquote {\bibinfo {title} {{Observation of radiation
  pressure shot noise on a macroscopic object}},}\ }\href
  {http://dx.doi.org/10.1126/science.1231282} {\bibfield  {journal} {\bibinfo
  {journal} {Science}\ }\textbf {\bibinfo {volume} {339}},\ \bibinfo {pages}
  {801} (\bibinfo {year} {2013}{\natexlab{a}})}\BibitemShut {NoStop}%
\bibitem [{\citenamefont {Nielsen}\ \emph {et~al.}(2017)\citenamefont
  {Nielsen}, \citenamefont {Tsaturyan}, \citenamefont {M{\o}ller},
  \citenamefont {Polzik},\ and\ \citenamefont
  {Schliesser}}]{Nielsen2017Multimode}%
  \BibitemOpen
  \bibfield  {author} {\bibinfo {author} {\bibfnamefont {W.~H.~P.}\
  \bibnamefont {Nielsen}}, \bibinfo {author} {\bibfnamefont {Y.}~\bibnamefont
  {Tsaturyan}}, \bibinfo {author} {\bibfnamefont {C.~B.}\ \bibnamefont
  {M{\o}ller}}, \bibinfo {author} {\bibfnamefont {E.~S.}\ \bibnamefont
  {Polzik}}, \ and\ \bibinfo {author} {\bibfnamefont {A.}~\bibnamefont
  {Schliesser}},\ }\bibfield  {title} {\enquote {\bibinfo {title} {{Multimode
  optomechanical system in the quantum regime}},}\ }\href
  {http://dx.doi.org/10.1073/pnas.1608412114} {\bibfield  {journal} {\bibinfo
  {journal} {Proceedings of the National Academy of Sciences}\ }\textbf
  {\bibinfo {volume} {114}},\ \bibinfo {pages} {62} (\bibinfo {year}
  {2017})}\BibitemShut {NoStop}%
\bibitem [{\citenamefont {Purdy}\ \emph
  {et~al.}(2013{\natexlab{b}})\citenamefont {Purdy}, \citenamefont {Yu},
  \citenamefont {Peterson}, \citenamefont {Kampel},\ and\ \citenamefont
  {Regal}}]{Purdy2013Strong}%
  \BibitemOpen
  \bibfield  {author} {\bibinfo {author} {\bibfnamefont {T.~P.}\ \bibnamefont
  {Purdy}}, \bibinfo {author} {\bibfnamefont {P.-L.~L.}\ \bibnamefont {Yu}},
  \bibinfo {author} {\bibfnamefont {R.~W.}\ \bibnamefont {Peterson}}, \bibinfo
  {author} {\bibfnamefont {N.~S.}\ \bibnamefont {Kampel}}, \ and\ \bibinfo
  {author} {\bibfnamefont {C.~A.}\ \bibnamefont {Regal}},\ }\bibfield  {title}
  {\enquote {\bibinfo {title} {{Strong optomechanical squeezing of light}},}\
  }\href {http://dx.doi.org/10.1103/PhysRevX.3.031012} {\bibfield  {journal}
  {\bibinfo  {journal} {Phys. Rev. X}\ }\textbf {\bibinfo {volume} {3}},\
  \bibinfo {pages} {031012} (\bibinfo {year} {2013}{\natexlab{b}})}\BibitemShut
  {NoStop}%
\bibitem [{\citenamefont {Rossi}\ \emph {et~al.}(2018)\citenamefont {Rossi},
  \citenamefont {Mason}, \citenamefont {Chen}, \citenamefont {Tsaturyan},\ and\
  \citenamefont {Schliesser}}]{Rossi2018Measurement}%
  \BibitemOpen
  \bibfield  {author} {\bibinfo {author} {\bibfnamefont {M.}~\bibnamefont
  {Rossi}}, \bibinfo {author} {\bibfnamefont {D.}~\bibnamefont {Mason}},
  \bibinfo {author} {\bibfnamefont {J.}~\bibnamefont {Chen}}, \bibinfo {author}
  {\bibfnamefont {Y.}~\bibnamefont {Tsaturyan}}, \ and\ \bibinfo {author}
  {\bibfnamefont {A.}~\bibnamefont {Schliesser}},\ }\bibfield  {title}
  {\enquote {\bibinfo {title} {{Measurement-based quantum control of mechanical
  motion}},}\ }\href {http://dx.doi.org/10.1038/s41586-018-0643-8} {\bibfield
  {journal} {\bibinfo  {journal} {Nature}\ }\textbf {\bibinfo {volume} {563}},\
  \bibinfo {pages} {53} (\bibinfo {year} {2018})}\BibitemShut {NoStop}%
\bibitem [{\citenamefont {Andrews}\ \emph {et~al.}(2014)\citenamefont
  {Andrews}, \citenamefont {Peterson}, \citenamefont {Purdy}, \citenamefont
  {Cicak}, \citenamefont {Simmonds}, \citenamefont {Regal},\ and\ \citenamefont
  {Lehnert}}]{Andrews2014Bidirectional}%
  \BibitemOpen
  \bibfield  {author} {\bibinfo {author} {\bibfnamefont {R.~W.}\ \bibnamefont
  {Andrews}}, \bibinfo {author} {\bibfnamefont {R.~W.}\ \bibnamefont
  {Peterson}}, \bibinfo {author} {\bibfnamefont {T.~P.}\ \bibnamefont {Purdy}},
  \bibinfo {author} {\bibfnamefont {K.}~\bibnamefont {Cicak}}, \bibinfo
  {author} {\bibfnamefont {R.~W.}\ \bibnamefont {Simmonds}}, \bibinfo {author}
  {\bibfnamefont {C.~A.}\ \bibnamefont {Regal}}, \ and\ \bibinfo {author}
  {\bibfnamefont {K.~W.}\ \bibnamefont {Lehnert}},\ }\bibfield  {title}
  {\enquote {\bibinfo {title} {{Bidirectional and efficient conversion between
  microwave and optical light}},}\ }\href {http://dx.doi.org/10.1038/nphys2911}
  {\bibfield  {journal} {\bibinfo  {journal} {Nat. Phys.}\ }\textbf {\bibinfo
  {volume} {10}},\ \bibinfo {pages} {321} (\bibinfo {year} {2014})}\BibitemShut
  {NoStop}%
\bibitem [{\citenamefont {Higginbotham}\ \emph {et~al.}(2018)\citenamefont
  {Higginbotham}, \citenamefont {Burns}, \citenamefont {Urmey}, \citenamefont
  {Peterson}, \citenamefont {Kampel}, \citenamefont {Brubaker}, \citenamefont
  {Smith}, \citenamefont {Lehnert},\ and\ \citenamefont
  {Regal}}]{Higginbotham2018Harnessing}%
  \BibitemOpen
  \bibfield  {author} {\bibinfo {author} {\bibfnamefont {A.~P.}\ \bibnamefont
  {Higginbotham}}, \bibinfo {author} {\bibfnamefont {P.~S.}\ \bibnamefont
  {Burns}}, \bibinfo {author} {\bibfnamefont {M.~D.}\ \bibnamefont {Urmey}},
  \bibinfo {author} {\bibfnamefont {R.~W.}\ \bibnamefont {Peterson}}, \bibinfo
  {author} {\bibfnamefont {N.~S.}\ \bibnamefont {Kampel}}, \bibinfo {author}
  {\bibfnamefont {B.~M.}\ \bibnamefont {Brubaker}}, \bibinfo {author}
  {\bibfnamefont {G.}~\bibnamefont {Smith}}, \bibinfo {author} {\bibfnamefont
  {K.~W.}\ \bibnamefont {Lehnert}}, \ and\ \bibinfo {author} {\bibfnamefont
  {C.~A.}\ \bibnamefont {Regal}},\ }\bibfield  {title} {\enquote {\bibinfo
  {title} {{Harnessing electro-optic correlations in an efficient mechanical
  converter}},}\ }\href {http://dx.doi.org/10.1038/s41567-018-0210-0}
  {\bibfield  {journal} {\bibinfo  {journal} {Nat. Phys.}\ }\textbf {\bibinfo
  {volume} {14}},\ \bibinfo {pages} {1038} (\bibinfo {year}
  {2018})}\BibitemShut {NoStop}%
\bibitem [{\citenamefont {Lee}\ \emph {et~al.}(2014)\citenamefont {Lee},
  \citenamefont {Underwood}, \citenamefont {Mason}, \citenamefont {Shkarin},
  \citenamefont {Hoch},\ and\ \citenamefont {Harris}}]{Lee2014Multimode}%
  \BibitemOpen
  \bibfield  {author} {\bibinfo {author} {\bibfnamefont {D.}~\bibnamefont
  {Lee}}, \bibinfo {author} {\bibfnamefont {M.}~\bibnamefont {Underwood}},
  \bibinfo {author} {\bibfnamefont {D.}~\bibnamefont {Mason}}, \bibinfo
  {author} {\bibfnamefont {A.~B.}\ \bibnamefont {Shkarin}}, \bibinfo {author}
  {\bibfnamefont {S.~W.}\ \bibnamefont {Hoch}}, \ and\ \bibinfo {author}
  {\bibfnamefont {J.~G.~E.}\ \bibnamefont {Harris}},\ }\bibfield  {title}
  {\enquote {\bibinfo {title} {{Multimode optomechanical dynamics in a cavity
  with avoided crossings}},}\ }\href {http://dx.doi.org/10.1038/ncomms7232}
  {\bibfield  {journal} {\bibinfo  {journal} {Nat. Comm.}\ }\textbf {\bibinfo
  {volume} {6}},\ \bibinfo {pages} {6232} (\bibinfo {year} {2014})}\BibitemShut
  {NoStop}%
\bibitem [{\citenamefont {Ni}\ \emph {et~al.}(2012)\citenamefont {Ni},
  \citenamefont {Norte}, \citenamefont {Wilson}, \citenamefont {Hood},
  \citenamefont {Chang}, \citenamefont {Painter},\ and\ \citenamefont
  {Kimble}}]{Ni2012Enhancement}%
  \BibitemOpen
  \bibfield  {author} {\bibinfo {author} {\bibfnamefont {K.~K.}\ \bibnamefont
  {Ni}}, \bibinfo {author} {\bibfnamefont {R.}~\bibnamefont {Norte}}, \bibinfo
  {author} {\bibfnamefont {D.~J.}\ \bibnamefont {Wilson}}, \bibinfo {author}
  {\bibfnamefont {J.~D.}\ \bibnamefont {Hood}}, \bibinfo {author}
  {\bibfnamefont {D.~E.}\ \bibnamefont {Chang}}, \bibinfo {author}
  {\bibfnamefont {O.}~\bibnamefont {Painter}}, \ and\ \bibinfo {author}
  {\bibfnamefont {H.~J.}\ \bibnamefont {Kimble}},\ }\bibfield  {title}
  {\enquote {\bibinfo {title} {{Enhancement of mechanical Q factors by optical
  trapping}},}\ }\href {http://dx.doi.org/10.1103/physrevlett.108.214302}
  {\bibfield  {journal} {\bibinfo  {journal} {Phys. Rev. Lett.}\ }\textbf
  {\bibinfo {volume} {108}},\ \bibinfo {pages} {214302} (\bibinfo {year}
  {2012})}\BibitemShut {NoStop}%
\bibitem [{\citenamefont {Chang}\ \emph {et~al.}(2012)\citenamefont {Chang},
  \citenamefont {Ni}, \citenamefont {Painter},\ and\ \citenamefont
  {Kimble}}]{Chang2012Ultrahigh}%
  \BibitemOpen
  \bibfield  {author} {\bibinfo {author} {\bibfnamefont {D.~E.}\ \bibnamefont
  {Chang}}, \bibinfo {author} {\bibfnamefont {K.-K.}\ \bibnamefont {Ni}},
  \bibinfo {author} {\bibfnamefont {O.}~\bibnamefont {Painter}}, \ and\
  \bibinfo {author} {\bibfnamefont {H.~J.}\ \bibnamefont {Kimble}},\ }\bibfield
   {title} {\enquote {\bibinfo {title} {{Ultrahigh- Q mechanical oscillators
  through optical trapping}},}\ }\href
  {http://dx.doi.org/10.1088/1367-2630/14/4/045002} {\bibfield  {journal}
  {\bibinfo  {journal} {New J. Phys.}\ }\textbf {\bibinfo {volume} {14}},\
  \bibinfo {pages} {45002} (\bibinfo {year} {2012})}\BibitemShut {NoStop}%
\bibitem [{\citenamefont {M{\"{u}}ller}\ \emph {et~al.}(2015)\citenamefont
  {M{\"{u}}ller}, \citenamefont {Reinhardt},\ and\ \citenamefont
  {Sankey}}]{Muller2015Enhanced}%
  \BibitemOpen
  \bibfield  {author} {\bibinfo {author} {\bibfnamefont {T.}~\bibnamefont
  {M{\"{u}}ller}}, \bibinfo {author} {\bibfnamefont {C.}~\bibnamefont
  {Reinhardt}}, \ and\ \bibinfo {author} {\bibfnamefont {J.}~\bibnamefont
  {Sankey}},\ }\bibfield  {title} {\enquote {\bibinfo {title} {{Enhanced
  optomechanical levitation of minimally supported dielectrics}},}\ }\href
  {http://dx.doi.org/10.1103/PhysRevA.91.053849} {\bibfield  {journal}
  {\bibinfo  {journal} {Phys. Rev. A}\ }\textbf {\bibinfo {volume} {91}},\
  \bibinfo {pages} {053849} (\bibinfo {year} {2015})}\BibitemShut {NoStop}%
\bibitem [{\citenamefont {Barasheed}\ \emph {et~al.}(2016)\citenamefont
  {Barasheed}, \citenamefont {M{\"{u}}ller},\ and\ \citenamefont
  {Sankey}}]{Barasheed2016Optically}%
  \BibitemOpen
  \bibfield  {author} {\bibinfo {author} {\bibfnamefont {A.}~\bibnamefont
  {Barasheed}}, \bibinfo {author} {\bibfnamefont {T.}~\bibnamefont
  {M{\"{u}}ller}}, \ and\ \bibinfo {author} {\bibfnamefont {J.}~\bibnamefont
  {Sankey}},\ }\bibfield  {title} {\enquote {\bibinfo {title} {{Optically
  defined mechanical geometry}},}\ }\href
  {http://dx.doi.org/10.1103/PhysRevA.93.053811} {\bibfield  {journal}
  {\bibinfo  {journal} {Phys. Rev. A}\ }\textbf {\bibinfo {volume} {93}},\
  \bibinfo {pages} {053811} (\bibinfo {year} {2016})}\BibitemShut {NoStop}%
\bibitem [{\citenamefont {Nunnenkamp}\ \emph {et~al.}(2010)\citenamefont
  {Nunnenkamp}, \citenamefont {B{\o}rkje}, \citenamefont {Harris},\ and\
  \citenamefont {Girvin}}]{Nunnenkamp2010Cooling}%
  \BibitemOpen
  \bibfield  {author} {\bibinfo {author} {\bibfnamefont {A.}~\bibnamefont
  {Nunnenkamp}}, \bibinfo {author} {\bibfnamefont {K.}~\bibnamefont
  {B{\o}rkje}}, \bibinfo {author} {\bibfnamefont {J.~G.~E.}\ \bibnamefont
  {Harris}}, \ and\ \bibinfo {author} {\bibfnamefont {S.~M.}\ \bibnamefont
  {Girvin}},\ }\bibfield  {title} {\enquote {\bibinfo {title} {{Cooling and
  squeezing via quadratic optomechanical coupling}},}\ }\href
  {http://dx.doi.org/10.1103/physreva.82.021806} {\bibfield  {journal}
  {\bibinfo  {journal} {Phys. Rev. A}\ }\textbf {\bibinfo {volume} {82}},\
  \bibinfo {pages} {021806} (\bibinfo {year} {2010})}\BibitemShut {NoStop}%
\bibitem [{\citenamefont {Jayich}\ \emph {et~al.}(2008)\citenamefont {Jayich},
  \citenamefont {Sankey}, \citenamefont {Zwickl}, \citenamefont {Yang},
  \citenamefont {Thompson}, \citenamefont {Girvin}, \citenamefont {Clerk},
  \citenamefont {Marquardt},\ and\ \citenamefont
  {Harris}}]{Jayich2008Dispersive}%
  \BibitemOpen
  \bibfield  {author} {\bibinfo {author} {\bibfnamefont {A.~M.}\ \bibnamefont
  {Jayich}}, \bibinfo {author} {\bibfnamefont {J.~C.}\ \bibnamefont {Sankey}},
  \bibinfo {author} {\bibfnamefont {B.~M.}\ \bibnamefont {Zwickl}}, \bibinfo
  {author} {\bibfnamefont {C.}~\bibnamefont {Yang}}, \bibinfo {author}
  {\bibfnamefont {J.~D.}\ \bibnamefont {Thompson}}, \bibinfo {author}
  {\bibfnamefont {S.~M.}\ \bibnamefont {Girvin}}, \bibinfo {author}
  {\bibfnamefont {A.~A.}\ \bibnamefont {Clerk}}, \bibinfo {author}
  {\bibfnamefont {F.}~\bibnamefont {Marquardt}}, \ and\ \bibinfo {author}
  {\bibfnamefont {J.~G.~E.}\ \bibnamefont {Harris}},\ }\bibfield  {title}
  {\enquote {\bibinfo {title} {{Dispersive optomechanics: a membrane inside a
  cavity}},}\ }\href {http://dx.doi.org/10.1088/1367-2630/10/9/095008}
  {\bibfield  {journal} {\bibinfo  {journal} {New J. Phys.}\ }\textbf {\bibinfo
  {volume} {10}},\ \bibinfo {pages} {095008} (\bibinfo {year}
  {2008})}\BibitemShut {NoStop}%
\bibitem [{\citenamefont {Miao}\ \emph {et~al.}(2009)\citenamefont {Miao},
  \citenamefont {Danilishin}, \citenamefont {Corbitt},\ and\ \citenamefont
  {Chen}}]{Miao2009Standard}%
  \BibitemOpen
  \bibfield  {author} {\bibinfo {author} {\bibfnamefont {H.}~\bibnamefont
  {Miao}}, \bibinfo {author} {\bibfnamefont {S.}~\bibnamefont {Danilishin}},
  \bibinfo {author} {\bibfnamefont {T.}~\bibnamefont {Corbitt}}, \ and\
  \bibinfo {author} {\bibfnamefont {Y.}~\bibnamefont {Chen}},\ }\bibfield
  {title} {\enquote {\bibinfo {title} {{Standard quantum limit for probing
  mechanical energy quantization}},}\ }\href
  {http://dx.doi.org/10.1103/PhysRevLett.103.100402} {\bibfield  {journal}
  {\bibinfo  {journal} {Phys. Rev. Lett.}\ }\textbf {\bibinfo {volume} {103}},\
  \bibinfo {pages} {100402} (\bibinfo {year} {2009})}\BibitemShut {NoStop}%
\bibitem [{\citenamefont {Yanay}\ \emph {et~al.}(2016)\citenamefont {Yanay},
  \citenamefont {Sankey},\ and\ \citenamefont {Clerk}}]{Yanay2016Quantum}%
  \BibitemOpen
  \bibfield  {author} {\bibinfo {author} {\bibfnamefont {Y.}~\bibnamefont
  {Yanay}}, \bibinfo {author} {\bibfnamefont {J.}~\bibnamefont {Sankey}}, \
  and\ \bibinfo {author} {\bibfnamefont {A.}~\bibnamefont {Clerk}},\ }\bibfield
   {title} {\enquote {\bibinfo {title} {{Quantum backaction and noise
  interference in asymmetric two-cavity optomechanical systems}},}\ }\href
  {http://dx.doi.org/10.1103/PhysRevA.93.063809} {\bibfield  {journal}
  {\bibinfo  {journal} {Phys. Rev. A}\ }\textbf {\bibinfo {volume} {93}},\
  \bibinfo {pages} {063809} (\bibinfo {year} {2016})}\BibitemShut {NoStop}%
\bibitem [{\citenamefont {Elste}\ \emph {et~al.}(2009)\citenamefont {Elste},
  \citenamefont {Girvin},\ and\ \citenamefont {Clerk}}]{Elste2009Quantum}%
  \BibitemOpen
  \bibfield  {author} {\bibinfo {author} {\bibfnamefont {F.}~\bibnamefont
  {Elste}}, \bibinfo {author} {\bibfnamefont {S.~M.}\ \bibnamefont {Girvin}}, \
  and\ \bibinfo {author} {\bibfnamefont {A.~A.}\ \bibnamefont {Clerk}},\
  }\bibfield  {title} {\enquote {\bibinfo {title} {{Quantum noise interference
  and backaction cooling in cavity nanomechanics}},}\ }\href
  {http://dx.doi.org/10.1103/PhysRevLett.102.207209} {\bibfield  {journal}
  {\bibinfo  {journal} {Phys. Rev. Lett.}\ }\textbf {\bibinfo {volume} {102}},\
  \bibinfo {pages} {207209} (\bibinfo {year} {2009})}\BibitemShut {NoStop}%
\bibitem [{\citenamefont {Kilda}\ and\ \citenamefont
  {Nunnenkamp}(2016)}]{Kilda2016Squeezed}%
  \BibitemOpen
  \bibfield  {author} {\bibinfo {author} {\bibfnamefont {D.}~\bibnamefont
  {Kilda}}\ and\ \bibinfo {author} {\bibfnamefont {A.}~\bibnamefont
  {Nunnenkamp}},\ }\bibfield  {title} {\enquote {\bibinfo {title} {{Squeezed
  light and correlated photons from dissipatively coupled optomechanical
  systems}},}\ }\href {http://dx.doi.org/10.1088/2040-8978/18/1/014007}
  {\bibfield  {journal} {\bibinfo  {journal} {J. Optics}\ }\textbf {\bibinfo
  {volume} {18}},\ \bibinfo {pages} {014007} (\bibinfo {year}
  {2016})}\BibitemShut {NoStop}%
\bibitem [{\citenamefont {Tarabrin}\ \emph {et~al.}(2013)\citenamefont
  {Tarabrin}, \citenamefont {Kaufer}, \citenamefont {Khalili}, \citenamefont
  {Schnabel},\ and\ \citenamefont {Hammerer}}]{Tarabrin2013Anomalous}%
  \BibitemOpen
  \bibfield  {author} {\bibinfo {author} {\bibfnamefont {S.~P.}\ \bibnamefont
  {Tarabrin}}, \bibinfo {author} {\bibfnamefont {H.}~\bibnamefont {Kaufer}},
  \bibinfo {author} {\bibfnamefont {F.~Y.}\ \bibnamefont {Khalili}}, \bibinfo
  {author} {\bibfnamefont {R.}~\bibnamefont {Schnabel}}, \ and\ \bibinfo
  {author} {\bibfnamefont {K.}~\bibnamefont {Hammerer}},\ }\bibfield  {title}
  {\enquote {\bibinfo {title} {{Anomalous dynamic backaction in
  interferometers}},}\ }\href {http://dx.doi.org/10.1103/PhysRevA.88.023809}
  {\bibfield  {journal} {\bibinfo  {journal} {Phys. Rev. A}\ }\textbf {\bibinfo
  {volume} {88}},\ \bibinfo {pages} {023809} (\bibinfo {year}
  {2013})}\BibitemShut {NoStop}%
\bibitem [{\citenamefont {Sawadsky}\ \emph {et~al.}(2015)\citenamefont
  {Sawadsky}, \citenamefont {Kaufer}, \citenamefont {Nia}, \citenamefont
  {Tarabrin}, \citenamefont {Khalili}, \citenamefont {Hammerer},\ and\
  \citenamefont {Schnabel}}]{Sawadsky2015Observation}%
  \BibitemOpen
  \bibfield  {author} {\bibinfo {author} {\bibfnamefont {A.}~\bibnamefont
  {Sawadsky}}, \bibinfo {author} {\bibfnamefont {H.}~\bibnamefont {Kaufer}},
  \bibinfo {author} {\bibfnamefont {R.~M.}\ \bibnamefont {Nia}}, \bibinfo
  {author} {\bibfnamefont {S.~P.}\ \bibnamefont {Tarabrin}}, \bibinfo {author}
  {\bibfnamefont {F.~Y.}\ \bibnamefont {Khalili}}, \bibinfo {author}
  {\bibfnamefont {K.}~\bibnamefont {Hammerer}}, \ and\ \bibinfo {author}
  {\bibfnamefont {R.}~\bibnamefont {Schnabel}},\ }\bibfield  {title} {\enquote
  {\bibinfo {title} {{Observation of generalized optomechanical coupling and
  cooling on cavity resonance}},}\ }\href
  {http://dx.doi.org/10.1103/PhysRevLett.114.043601} {\bibfield  {journal}
  {\bibinfo  {journal} {Phys. Rev. Lett.}\ }\textbf {\bibinfo {volume} {114}},\
  \bibinfo {pages} {043601} (\bibinfo {year} {2015})}\BibitemShut {NoStop}%
\bibitem [{\citenamefont {Flowers-Jacobs}\ \emph {et~al.}(2012)\citenamefont
  {Flowers-Jacobs}, \citenamefont {Hoch}, \citenamefont {Sankey}, \citenamefont
  {Kashkanova}, \citenamefont {Jayich}, \citenamefont {Deutsch}, \citenamefont
  {Reichel},\ and\ \citenamefont {Harris}}]{Flowers2012Fiber}%
  \BibitemOpen
  \bibfield  {author} {\bibinfo {author} {\bibfnamefont {N.}~\bibnamefont
  {Flowers-Jacobs}}, \bibinfo {author} {\bibfnamefont {S.}~\bibnamefont
  {Hoch}}, \bibinfo {author} {\bibfnamefont {J.}~\bibnamefont {Sankey}},
  \bibinfo {author} {\bibfnamefont {A.}~\bibnamefont {Kashkanova}}, \bibinfo
  {author} {\bibfnamefont {A.}~\bibnamefont {Jayich}}, \bibinfo {author}
  {\bibfnamefont {C.}~\bibnamefont {Deutsch}}, \bibinfo {author} {\bibfnamefont
  {J.}~\bibnamefont {Reichel}}, \ and\ \bibinfo {author} {\bibfnamefont
  {J.}~\bibnamefont {Harris}},\ }\bibfield  {title} {\enquote {\bibinfo {title}
  {{Fiber-cavity-based optomechanical device}},}\ }\href
  {http://dx.doi.org/10.1063/1.4768779} {\bibfield  {journal} {\bibinfo
  {journal} {Appl. Phys. Lett.}\ }\textbf {\bibinfo {volume} {101}},\ \bibinfo
  {pages} {221109} (\bibinfo {year} {2012})}\BibitemShut {NoStop}%
\bibitem [{\citenamefont {Arrangoiz-Arriola}\ \emph {et~al.}(2019)\citenamefont
  {Arrangoiz-Arriola}, \citenamefont {Wollack}, \citenamefont {Wang},
  \citenamefont {Pechal}, \citenamefont {Jiang}, \citenamefont {McKenna},
  \citenamefont {Witmer},\ and\ \citenamefont
  {Safavi-Naeini}}]{Arrangoiz2019ResolvingARXIV}%
  \BibitemOpen
  \bibfield  {author} {\bibinfo {author} {\bibfnamefont {P.}~\bibnamefont
  {Arrangoiz-Arriola}}, \bibinfo {author} {\bibfnamefont {E.~A.}\ \bibnamefont
  {Wollack}}, \bibinfo {author} {\bibfnamefont {Z.}~\bibnamefont {Wang}},
  \bibinfo {author} {\bibfnamefont {M.}~\bibnamefont {Pechal}}, \bibinfo
  {author} {\bibfnamefont {W.}~\bibnamefont {Jiang}}, \bibinfo {author}
  {\bibfnamefont {T.~P.}\ \bibnamefont {McKenna}}, \bibinfo {author}
  {\bibfnamefont {J.~D.}\ \bibnamefont {Witmer}}, \ and\ \bibinfo {author}
  {\bibfnamefont {A.~H.}\ \bibnamefont {Safavi-Naeini}},\ }\bibfield  {title}
  {\enquote {\bibinfo {title} {{Resolving the energy levels of a nanomechanical
  oscillator}},}\ }\href {http://arxiv.org/abs/1902.04681} {\bibfield
  {journal} {\bibinfo  {journal} {arXiv:1902.04681}\ } (\bibinfo {year}
  {2019})}\BibitemShut {NoStop}%
\bibitem [{\citenamefont {Meenehan}\ \emph {et~al.}(2015)\citenamefont
  {Meenehan}, \citenamefont {Cohen}, \citenamefont {MacCabe}, \citenamefont
  {Marsili}, \citenamefont {Shaw},\ and\ \citenamefont
  {Painter}}]{Meenehan2015Pulsed}%
  \BibitemOpen
  \bibfield  {author} {\bibinfo {author} {\bibfnamefont {S.~M.}\ \bibnamefont
  {Meenehan}}, \bibinfo {author} {\bibfnamefont {J.~D.}\ \bibnamefont {Cohen}},
  \bibinfo {author} {\bibfnamefont {G.~S.}\ \bibnamefont {MacCabe}}, \bibinfo
  {author} {\bibfnamefont {F.}~\bibnamefont {Marsili}}, \bibinfo {author}
  {\bibfnamefont {M.~D.}\ \bibnamefont {Shaw}}, \ and\ \bibinfo {author}
  {\bibfnamefont {O.}~\bibnamefont {Painter}},\ }\bibfield  {title} {\enquote
  {\bibinfo {title} {{Pulsed excitation dynamics of an optomechanical crystal
  resonator near its quantum ground state of motion}},}\ }\href
  {http://dx.doi.org/10.1103/PhysRevX.5.041002} {\bibfield  {journal} {\bibinfo
   {journal} {Phys. Rev. X}\ }\textbf {\bibinfo {volume} {5}},\ \bibinfo
  {pages} {041002} (\bibinfo {year} {2015})}\BibitemShut {NoStop}%
\bibitem [{\citenamefont {Marinkovi{\'{c}}}\ \emph {et~al.}(2018)\citenamefont
  {Marinkovi{\'{c}}}, \citenamefont {Wallucks}, \citenamefont {Riedinger},
  \citenamefont {Hong}, \citenamefont {Aspelmeyer},\ and\ \citenamefont
  {Gr{\"{o}}blacher}}]{Marinkovic2018Optomechanical}%
  \BibitemOpen
  \bibfield  {author} {\bibinfo {author} {\bibfnamefont {I.}~\bibnamefont
  {Marinkovi{\'{c}}}}, \bibinfo {author} {\bibfnamefont {A.}~\bibnamefont
  {Wallucks}}, \bibinfo {author} {\bibfnamefont {R.}~\bibnamefont {Riedinger}},
  \bibinfo {author} {\bibfnamefont {S.}~\bibnamefont {Hong}}, \bibinfo {author}
  {\bibfnamefont {M.}~\bibnamefont {Aspelmeyer}}, \ and\ \bibinfo {author}
  {\bibfnamefont {S.}~\bibnamefont {Gr{\"{o}}blacher}},\ }\bibfield  {title}
  {\enquote {\bibinfo {title} {{Optomechanical Bell test}},}\ }\href
  {http://dx.doi.org/10.1103/PhysRevLett.121.220404} {\bibfield  {journal}
  {\bibinfo  {journal} {Phys. Rev. Lett.}\ }\textbf {\bibinfo {volume} {121}},\
  \bibinfo {pages} {220404} (\bibinfo {year} {2018})}\BibitemShut {NoStop}%
\bibitem [{\citenamefont {Riedinger}\ \emph {et~al.}(2018)\citenamefont
  {Riedinger}, \citenamefont {Wallucks}, \citenamefont {Marinkovi{\'{c}}},
  \citenamefont {L{\"{o}}schnauer}, \citenamefont {Aspelmeyer}, \citenamefont
  {Hong},\ and\ \citenamefont {Gr{\"{o}}blacher}}]{Riedinger2018Remote}%
  \BibitemOpen
  \bibfield  {author} {\bibinfo {author} {\bibfnamefont {R.}~\bibnamefont
  {Riedinger}}, \bibinfo {author} {\bibfnamefont {A.}~\bibnamefont {Wallucks}},
  \bibinfo {author} {\bibfnamefont {I.}~\bibnamefont {Marinkovi{\'{c}}}},
  \bibinfo {author} {\bibfnamefont {C.}~\bibnamefont {L{\"{o}}schnauer}},
  \bibinfo {author} {\bibfnamefont {M.}~\bibnamefont {Aspelmeyer}}, \bibinfo
  {author} {\bibfnamefont {S.}~\bibnamefont {Hong}}, \ and\ \bibinfo {author}
  {\bibfnamefont {S.}~\bibnamefont {Gr{\"{o}}blacher}},\ }\bibfield  {title}
  {\enquote {\bibinfo {title} {{Remote quantum entanglement between two
  micromechanical oscillators}},}\ }\href
  {http://dx.doi.org/10.1038/s41586-018-0036-z} {\bibfield  {journal} {\bibinfo
   {journal} {Nature}\ }\textbf {\bibinfo {volume} {556}},\ \bibinfo {pages}
  {473} (\bibinfo {year} {2018})}\BibitemShut {NoStop}%
\bibitem [{\citenamefont {Hong}\ \emph {et~al.}(2017)\citenamefont {Hong},
  \citenamefont {Riedinger}, \citenamefont {Marinkovi{\'{c}}}, \citenamefont
  {Wallucks}, \citenamefont {Hofer}, \citenamefont {Norte}, \citenamefont
  {Aspelmeyer},\ and\ \citenamefont {Gr{\"{o}}blacher}}]{Hong2017Hanbury}%
  \BibitemOpen
  \bibfield  {author} {\bibinfo {author} {\bibfnamefont {S.}~\bibnamefont
  {Hong}}, \bibinfo {author} {\bibfnamefont {R.}~\bibnamefont {Riedinger}},
  \bibinfo {author} {\bibfnamefont {I.}~\bibnamefont {Marinkovi{\'{c}}}},
  \bibinfo {author} {\bibfnamefont {A.}~\bibnamefont {Wallucks}}, \bibinfo
  {author} {\bibfnamefont {S.~G.}\ \bibnamefont {Hofer}}, \bibinfo {author}
  {\bibfnamefont {R.~A.}\ \bibnamefont {Norte}}, \bibinfo {author}
  {\bibfnamefont {M.}~\bibnamefont {Aspelmeyer}}, \ and\ \bibinfo {author}
  {\bibfnamefont {S.}~\bibnamefont {Gr{\"{o}}blacher}},\ }\bibfield  {title}
  {\enquote {\bibinfo {title} {{Hanbury Brown and Twiss interferometry of
  single phonons from an optomechanical resonator.}}}\ }\href
  {http://dx.doi.org/10.1126/science.aan7939} {\bibfield  {journal} {\bibinfo
  {journal} {Science}\ }\textbf {\bibinfo {volume} {358}},\ \bibinfo {pages}
  {203} (\bibinfo {year} {2017})}\BibitemShut {NoStop}%
\bibitem [{\citenamefont {Nair}\ \emph {et~al.}(2017)\citenamefont {Nair},
  \citenamefont {Naesby},\ and\ \citenamefont
  {Dantan}}]{Nair2017Optomechanical}%
  \BibitemOpen
  \bibfield  {author} {\bibinfo {author} {\bibfnamefont {B.}~\bibnamefont
  {Nair}}, \bibinfo {author} {\bibfnamefont {A.}~\bibnamefont {Naesby}}, \ and\
  \bibinfo {author} {\bibfnamefont {A.}~\bibnamefont {Dantan}},\ }\bibfield
  {title} {\enquote {\bibinfo {title} {{Optomechanical characterization of
  silicon nitride membrane arrays}},}\ }\href
  {http://dx.doi.org/10.1364/OL.42.001341} {\bibfield  {journal} {\bibinfo
  {journal} {Optics Lett.}\ }\textbf {\bibinfo {volume} {42}},\ \bibinfo
  {pages} {1341} (\bibinfo {year} {2017})}\BibitemShut {NoStop}%
\bibitem [{\citenamefont {Piergentili}\ \emph {et~al.}(2018)\citenamefont
  {Piergentili}, \citenamefont {Catalini}, \citenamefont {Bawaj}, \citenamefont
  {Zippilli}, \citenamefont {Malossi}, \citenamefont {Natali}, \citenamefont
  {Vitali},\ and\ \citenamefont {Giuseppe}}]{Piergentili2018Two}%
  \BibitemOpen
  \bibfield  {author} {\bibinfo {author} {\bibfnamefont {P.}~\bibnamefont
  {Piergentili}}, \bibinfo {author} {\bibfnamefont {L.}~\bibnamefont
  {Catalini}}, \bibinfo {author} {\bibfnamefont {M.}~\bibnamefont {Bawaj}},
  \bibinfo {author} {\bibfnamefont {S.}~\bibnamefont {Zippilli}}, \bibinfo
  {author} {\bibfnamefont {N.}~\bibnamefont {Malossi}}, \bibinfo {author}
  {\bibfnamefont {R.}~\bibnamefont {Natali}}, \bibinfo {author} {\bibfnamefont
  {D.}~\bibnamefont {Vitali}}, \ and\ \bibinfo {author} {\bibfnamefont {G.~D.}\
  \bibnamefont {Giuseppe}},\ }\bibfield  {title} {\enquote {\bibinfo {title}
  {{Two-membrane cavity optomechanics}},}\ }\href
  {http://dx.doi.org/10.1088/1367-2630/aad85f} {\bibfield  {journal} {\bibinfo
  {journal} {New J. Phys.}\ }\textbf {\bibinfo {volume} {20}},\ \bibinfo
  {pages} {083024} (\bibinfo {year} {2018})}\BibitemShut {NoStop}%
\bibitem [{\citenamefont {Li}\ \emph {et~al.}(2016)\citenamefont {Li},
  \citenamefont {Xuereb}, \citenamefont {Malossi},\ and\ \citenamefont
  {Vitali}}]{Li2016Cavity}%
  \BibitemOpen
  \bibfield  {author} {\bibinfo {author} {\bibfnamefont {J.}~\bibnamefont
  {Li}}, \bibinfo {author} {\bibfnamefont {A.}~\bibnamefont {Xuereb}}, \bibinfo
  {author} {\bibfnamefont {N.}~\bibnamefont {Malossi}}, \ and\ \bibinfo
  {author} {\bibfnamefont {D.}~\bibnamefont {Vitali}},\ }\bibfield  {title}
  {\enquote {\bibinfo {title} {{Cavity mode frequencies and strong
  optomechanical coupling in two-membrane cavity optomechanics}},}\ }\href
  {http://dx.doi.org/10.1088/2040-8978/18/8/084001} {\bibfield  {journal}
  {\bibinfo  {journal} {J. Optics}\ }\textbf {\bibinfo {volume} {18}},\
  \bibinfo {pages} {084001} (\bibinfo {year} {2016})}\BibitemShut {NoStop}%
\bibitem [{\citenamefont {Leslie}\ \emph {et~al.}(2010)\citenamefont {Leslie},
  \citenamefont {Fields},\ and\ \citenamefont {Cohen}}]{Leslie2010Convex}%
  \BibitemOpen
  \bibfield  {author} {\bibinfo {author} {\bibfnamefont {S.~R.}\ \bibnamefont
  {Leslie}}, \bibinfo {author} {\bibfnamefont {A.~P.}\ \bibnamefont {Fields}},
  \ and\ \bibinfo {author} {\bibfnamefont {A.~E.}\ \bibnamefont {Cohen}},\
  }\bibfield  {title} {\enquote {\bibinfo {title} {{Convex lens-induced
  confinement for imaging single molecules}},}\ }\href
  {http://dx.doi.org/10.1021/ac101041s} {\bibfield  {journal} {\bibinfo
  {journal} {Analytical Chem.}\ }\textbf {\bibinfo {volume} {82}},\ \bibinfo
  {pages} {6224} (\bibinfo {year} {2010})}\BibitemShut {NoStop}%
\bibitem [{\citenamefont {Stambaugh}\ \emph {et~al.}(2015)\citenamefont
  {Stambaugh}, \citenamefont {Xu}, \citenamefont {Kemiktarak}, \citenamefont
  {Taylor},\ and\ \citenamefont {Lawall}}]{Stambaugh2015From}%
  \BibitemOpen
  \bibfield  {author} {\bibinfo {author} {\bibfnamefont {C.}~\bibnamefont
  {Stambaugh}}, \bibinfo {author} {\bibfnamefont {H.}~\bibnamefont {Xu}},
  \bibinfo {author} {\bibfnamefont {U.}~\bibnamefont {Kemiktarak}}, \bibinfo
  {author} {\bibfnamefont {J.}~\bibnamefont {Taylor}}, \ and\ \bibinfo {author}
  {\bibfnamefont {J.}~\bibnamefont {Lawall}},\ }\bibfield  {title} {\enquote
  {\bibinfo {title} {{From membrane-in-the-middle to mirror-in-the-middle with
  a high-reflectivity sub-wavelength grating}},}\ }\href
  {http://dx.doi.org/10.1002/andp.201400142} {\bibfield  {journal} {\bibinfo
  {journal} {Annalen der Physik}\ }\textbf {\bibinfo {volume} {527}},\ \bibinfo
  {pages} {81} (\bibinfo {year} {2015})}\BibitemShut {NoStop}%
\bibitem [{\citenamefont {Chen}\ \emph {et~al.}(2017)\citenamefont {Chen},
  \citenamefont {Chardin}, \citenamefont {Makles}, \citenamefont {Ca{\"{e}}r},
  \citenamefont {Chua}, \citenamefont {Braive}, \citenamefont {Robert-Philip},
  \citenamefont {Briant}, \citenamefont {Cohadon}, \citenamefont {Heidmann},
  \citenamefont {Jacqmin},\ and\ \citenamefont
  {Del{\'{e}}glise}}]{Chen2017High}%
  \BibitemOpen
  \bibfield  {author} {\bibinfo {author} {\bibfnamefont {X.}~\bibnamefont
  {Chen}}, \bibinfo {author} {\bibfnamefont {C.}~\bibnamefont {Chardin}},
  \bibinfo {author} {\bibfnamefont {K.}~\bibnamefont {Makles}}, \bibinfo
  {author} {\bibfnamefont {C.}~\bibnamefont {Ca{\"{e}}r}}, \bibinfo {author}
  {\bibfnamefont {S.}~\bibnamefont {Chua}}, \bibinfo {author} {\bibfnamefont
  {R.}~\bibnamefont {Braive}}, \bibinfo {author} {\bibfnamefont
  {I.}~\bibnamefont {Robert-Philip}}, \bibinfo {author} {\bibfnamefont
  {T.}~\bibnamefont {Briant}}, \bibinfo {author} {\bibfnamefont {P.-F.}\
  \bibnamefont {Cohadon}}, \bibinfo {author} {\bibfnamefont {A.}~\bibnamefont
  {Heidmann}}, \bibinfo {author} {\bibfnamefont {T.}~\bibnamefont {Jacqmin}}, \
  and\ \bibinfo {author} {\bibfnamefont {S.}~\bibnamefont {Del{\'{e}}glise}},\
  }\bibfield  {title} {\enquote {\bibinfo {title} {{High-finesse Fabry-Perot
  cavities with bidimensional Si3N4 photonic-crystal slabs}},}\ }\href
  {http://dx.doi.org/10.1038/lsa.2016.190} {\bibfield  {journal} {\bibinfo
  {journal} {Light: Science {\&} Applications}\ }\textbf {\bibinfo {volume}
  {6}},\ \bibinfo {pages} {e16190} (\bibinfo {year} {2017})}\BibitemShut
  {NoStop}%
\bibitem [{\citenamefont {Paraiso}\ \emph {et~al.}(2015)\citenamefont
  {Paraiso}, \citenamefont {Kalaee}, \citenamefont {Zang}, \citenamefont
  {Pfeifer}, \citenamefont {Marquardt},\ and\ \citenamefont
  {Painter}}]{Paraiso2015Position}%
  \BibitemOpen
  \bibfield  {author} {\bibinfo {author} {\bibfnamefont {T.~K.}\ \bibnamefont
  {Paraiso}}, \bibinfo {author} {\bibfnamefont {M.}~\bibnamefont {Kalaee}},
  \bibinfo {author} {\bibfnamefont {L.}~\bibnamefont {Zang}}, \bibinfo {author}
  {\bibfnamefont {H.}~\bibnamefont {Pfeifer}}, \bibinfo {author} {\bibfnamefont
  {F.}~\bibnamefont {Marquardt}}, \ and\ \bibinfo {author} {\bibfnamefont
  {O.}~\bibnamefont {Painter}},\ }\bibfield  {title} {\enquote {\bibinfo
  {title} {{Position-squared coupling in a tunable photonic crystal
  optomechanical cavity}},}\ }\href
  {http://dx.doi.org/10.1103/PhysRevX.5.041024} {\bibfield  {journal} {\bibinfo
   {journal} {Phys. Rev. X}\ }\textbf {\bibinfo {volume} {5}},\ \bibinfo
  {pages} {041024} (\bibinfo {year} {2015})}\BibitemShut {NoStop}%
\bibitem [{\citenamefont {Hood}\ \emph {et~al.}(2001)\citenamefont {Hood},
  \citenamefont {Kimble},\ and\ \citenamefont {Ye}}]{Hood2001Characterization}%
  \BibitemOpen
  \bibfield  {author} {\bibinfo {author} {\bibfnamefont {C.}~\bibnamefont
  {Hood}}, \bibinfo {author} {\bibfnamefont {H.}~\bibnamefont {Kimble}}, \ and\
  \bibinfo {author} {\bibfnamefont {J.}~\bibnamefont {Ye}},\ }\bibfield
  {title} {\enquote {\bibinfo {title} {{Characterization of high-finesse
  mirrors: Loss, phase shifts, and mode structure in an optical cavity}},}\
  }\href {http://dx.doi.org/10.1103/PhysRevA.64.033804} {\bibfield  {journal}
  {\bibinfo  {journal} {Phys. Rev. A}\ }\textbf {\bibinfo {volume} {64}},\
  \bibinfo {pages} {033804} (\bibinfo {year} {2001})}\BibitemShut {NoStop}%
\bibitem [{\citenamefont {Sankey}\ \emph
  {et~al.}(2010{\natexlab{b}})\citenamefont {Sankey}, \citenamefont {Yang},
  \citenamefont {Zwickl}, \citenamefont {Jayich},\ and\ \citenamefont
  {Harris}}]{Sankey2010Nonlinear}%
  \BibitemOpen
  \bibfield  {author} {\bibinfo {author} {\bibfnamefont {J.}~\bibnamefont
  {Sankey}}, \bibinfo {author} {\bibfnamefont {C.}~\bibnamefont {Yang}},
  \bibinfo {author} {\bibfnamefont {B.}~\bibnamefont {Zwickl}}, \bibinfo
  {author} {\bibfnamefont {A.}~\bibnamefont {Jayich}}, \ and\ \bibinfo {author}
  {\bibfnamefont {J.}~\bibnamefont {Harris}},\ }\bibfield  {title} {\enquote
  {\bibinfo {title} {{Nonlinear optomechanical couplings: Tools for dealing
  with solid mechanical objects in the quantum regime}},}\ }in\ \href@noop {}
  {\emph {\bibinfo {booktitle} {Optics InfoBase Conference Papers}}}\ (\bibinfo
  {year} {2010})\BibitemShut {NoStop}%
\bibitem [{\citenamefont {Karuza}\ \emph {et~al.}(2011)\citenamefont {Karuza},
  \citenamefont {Galassi}, \citenamefont {Biancofiore}, \citenamefont
  {Molinelli}, \citenamefont {Natali}, \citenamefont {Tombesi}, \citenamefont
  {{Di Giuseppe}},\ and\ \citenamefont {Vitali}}]{Karuza2011Tunable}%
  \BibitemOpen
  \bibfield  {author} {\bibinfo {author} {\bibfnamefont {M.}~\bibnamefont
  {Karuza}}, \bibinfo {author} {\bibfnamefont {M.}~\bibnamefont {Galassi}},
  \bibinfo {author} {\bibfnamefont {C.}~\bibnamefont {Biancofiore}}, \bibinfo
  {author} {\bibfnamefont {C.}~\bibnamefont {Molinelli}}, \bibinfo {author}
  {\bibfnamefont {R.}~\bibnamefont {Natali}}, \bibinfo {author} {\bibfnamefont
  {P.}~\bibnamefont {Tombesi}}, \bibinfo {author} {\bibfnamefont
  {G.}~\bibnamefont {{Di Giuseppe}}}, \ and\ \bibinfo {author} {\bibfnamefont
  {D.}~\bibnamefont {Vitali}},\ }\bibfield  {title} {\enquote {\bibinfo {title}
  {{Tunable linear and quadratic optomechanical coupling for a tilted membrane
  within an optical cavity: theory and experiment}},}\ }\href
  {http://dx.doi.org/10.1088/2040-8978/15/2/025704} {\bibfield  {journal}
  {\bibinfo  {journal} {J. Optics}\ }\textbf {\bibinfo {volume} {15}},\
  \bibinfo {pages} {025704} (\bibinfo {year} {2011})}\BibitemShut {NoStop}%
\bibitem [{\citenamefont {Hunger}\ \emph {et~al.}(2010)\citenamefont {Hunger},
  \citenamefont {Steinmetz}, \citenamefont {Colombe}, \citenamefont {Deutsch},
  \citenamefont {H{\"{a}}nsch},\ and\ \citenamefont {Reichel}}]{Hunger2010A}%
  \BibitemOpen
  \bibfield  {author} {\bibinfo {author} {\bibfnamefont {D.}~\bibnamefont
  {Hunger}}, \bibinfo {author} {\bibfnamefont {T.}~\bibnamefont {Steinmetz}},
  \bibinfo {author} {\bibfnamefont {Y.}~\bibnamefont {Colombe}}, \bibinfo
  {author} {\bibfnamefont {C.}~\bibnamefont {Deutsch}}, \bibinfo {author}
  {\bibfnamefont {T.~W.}\ \bibnamefont {H{\"{a}}nsch}}, \ and\ \bibinfo
  {author} {\bibfnamefont {J.}~\bibnamefont {Reichel}},\ }\bibfield  {title}
  {\enquote {\bibinfo {title} {{A fiber Fabry-Perot cavity with high
  finesse}},}\ }\href {http://dx.doi.org/10.1088/1367-2630/12/6/065038}
  {\bibfield  {journal} {\bibinfo  {journal} {New J. Phys.}\ }\textbf {\bibinfo
  {volume} {12}},\ \bibinfo {pages} {065038} (\bibinfo {year}
  {2010})}\BibitemShut {NoStop}%
\bibitem [{\citenamefont {Sankey}\ \emph {et~al.}(2008)\citenamefont {Sankey},
  \citenamefont {Jayich}, \citenamefont {Zwickl}, \citenamefont {Yang},\ and\
  \citenamefont {Harris}}]{Sankey2008Improved}%
  \BibitemOpen
  \bibfield  {author} {\bibinfo {author} {\bibfnamefont {J.~C.}\ \bibnamefont
  {Sankey}}, \bibinfo {author} {\bibfnamefont {A.~M.}\ \bibnamefont {Jayich}},
  \bibinfo {author} {\bibfnamefont {B.~M.}\ \bibnamefont {Zwickl}}, \bibinfo
  {author} {\bibfnamefont {C.}~\bibnamefont {Yang}}, \ and\ \bibinfo {author}
  {\bibfnamefont {J.~G.~E.}\ \bibnamefont {Harris}},\ }\bibfield  {title}
  {\enquote {\bibinfo {title} {{Improved "position-squared" readout using
  degenerate cavity modes}},}\ }in\ \href
  {http://www.phys.uconn.edu/icap2008/proceedings/} {\emph {\bibinfo
  {booktitle} {ICAP Proceedings}}}\ (\bibinfo {year} {2008})\ pp.\ \bibinfo
  {pages} {131--149}\BibitemShut {NoStop}%
\bibitem [{\citenamefont {Weiss}\ \emph {et~al.}(2013)\citenamefont {Weiss},
  \citenamefont {Bruder},\ and\ \citenamefont {Nunnenkamp}}]{Weiss2013Strong}%
  \BibitemOpen
  \bibfield  {author} {\bibinfo {author} {\bibfnamefont {T.}~\bibnamefont
  {Weiss}}, \bibinfo {author} {\bibfnamefont {C.}~\bibnamefont {Bruder}}, \
  and\ \bibinfo {author} {\bibfnamefont {A.}~\bibnamefont {Nunnenkamp}},\
  }\bibfield  {title} {\enquote {\bibinfo {title} {{Strong-coupling effects in
  dissipatively coupled optomechanical systems}},}\ }\href
  {http://dx.doi.org/10.1088/1367-2630/15/4/045017} {\bibfield  {journal}
  {\bibinfo  {journal} {New J. Phys.}\ }\textbf {\bibinfo {volume} {15}},\
  \bibinfo {pages} {045017} (\bibinfo {year} {2013})}\BibitemShut {NoStop}%
\bibitem [{\citenamefont {Weiss}\ and\ \citenamefont
  {Nunnenkamp}(2013)}]{Weiss2013Quantum}%
  \BibitemOpen
  \bibfield  {author} {\bibinfo {author} {\bibfnamefont {T.}~\bibnamefont
  {Weiss}}\ and\ \bibinfo {author} {\bibfnamefont {A.}~\bibnamefont
  {Nunnenkamp}},\ }\bibfield  {title} {\enquote {\bibinfo {title} {{Quantum
  limit of laser cooling in dispersively and dissipatively coupled
  optomechanical systems}},}\ }\href
  {http://dx.doi.org/10.1103/PhysRevA.88.023850} {\bibfield  {journal}
  {\bibinfo  {journal} {Phys. Rev. A}\ }\textbf {\bibinfo {volume} {88}},\
  \bibinfo {pages} {023850} (\bibinfo {year} {2013})}\BibitemShut {NoStop}%
\bibitem [{\citenamefont {Purdy}\ \emph {et~al.}(2012)\citenamefont {Purdy},
  \citenamefont {Peterson}, \citenamefont {Yu},\ and\ \citenamefont
  {Regal}}]{Purdy2012Cavity}%
  \BibitemOpen
  \bibfield  {author} {\bibinfo {author} {\bibfnamefont {T.~P.}\ \bibnamefont
  {Purdy}}, \bibinfo {author} {\bibfnamefont {R.~W.}\ \bibnamefont {Peterson}},
  \bibinfo {author} {\bibfnamefont {P.~L.}\ \bibnamefont {Yu}}, \ and\ \bibinfo
  {author} {\bibfnamefont {C.~A.}\ \bibnamefont {Regal}},\ }\bibfield  {title}
  {\enquote {\bibinfo {title} {{Cavity optomechanics with Si3N4 membranes at
  cryogenic temperatures}},}\ }\href
  {http://dx.doi.org/10.1088/1367-2630/14/11/115021} {\bibfield  {journal}
  {\bibinfo  {journal} {New J. Phys.}\ }\textbf {\bibinfo {volume} {14}},\
  \bibinfo {pages} {115021} (\bibinfo {year} {2012})}\BibitemShut {NoStop}%
\end{thebibliography}%

\end{document}